\documentclass[preprint]{aastex63}
\usepackage{xspace}
\usepackage{amsmath}
\usepackage{booktabs}
\usepackage{longtable}
\usepackage{morefloats}
\usepackage{blindtext}
\usepackage{placeins}
\usepackage{makecell}
\usepackage[shortlabels]{enumitem}
\usepackage{rotating} 
\usepackage{graphicx}
\usepackage{changepage}
\usepackage[caption=false]{subfig}

\usepackage{color}

\newcommand{\Fermi}{\emph{Fermi}\xspace}
\newcommand{\lat}{\emph{Fermi}-LAT\xspace}
\newcommand{\Swift}{\emph{Swift}\xspace}

\begin{document}


 

\title{On the Existence of the Plateau Emission in High-Energy Gamma-Ray Burst Light Curves observed by \Fermi-LAT}
\author{M.G. Dainotti}
\affiliation{National Astronomical Observatory of Japan, 2-21-1 Osawa, Mitaka, Tokyo 181-8588,}
\affiliation{Space Science Institute, Boulder, Colorado}
\email{mdainott@stanford.edu}
\author{N. Omodei} 
\email{nicola.omodei@stanford.edu}
\affiliation{Physics Department, Stanford University, 382 Via Pueblo Mall, Stanford, USA}
\author{G.P. Srinivasaragavan}
\affiliation{Cahill Center for Astrophysics, California Institute of Technology, 1200 E. California Blvd. Pasadena, CA 91125, USA}
\email{gsriniva@caltech.edu, First 3 authors share same contribution}
\author{G. Vianello}
\affiliation{Physics Department, Stanford University, 382 Via Pueblo Mall, Stanford, USA}
\author{R. Willingale}
\affiliation{School of Physics and Astronomy, University of Leicester, UK}
\author{P. O'Brien}    
\affiliation{Astronomy Department, University of Leicester, UK}
\author{S. Nagataki}
\affiliation{RIKEN Cluster for Pioneering Research, Astrophysical Big Bang Laboratory (ABBL)}
\author{V. Petrosian}
\affiliation{Physics Department, Stanford University, 382 Via Pueblo Mall, Stanford, USA}
\author{Z. Nuygen}
\affiliation{Astronomy Department, UCLA, 475 Portola Plaza, Los Angeles, USA} 
\author{X. Hernandez}
\affiliation{Universidad Nacional Aut{\'o}noma de M{\'e}xico, Instituto de Astronom{\'{\i}}a, AP 70-264, CDMX  04510, México}, 
\author{M. Axelsson}
\affiliation{Department of Physics and Department of Astronomy, Stockholm University, 106 91 Stockholm, Sweden}
\affiliation{Department of Physics, KTH Royal Institute of Technology, and The Oskar Klein Centre, 106 91 Stockholm, Sweden} 
\author{E. Bissaldi}
\affiliation{Dipartimento Interateneo di Fisica dell'Universit{\`a} e Politecnico di Bari - Via E. Orabona 4, 70125 Bari, Italy}
\affiliation{INFN Sezione di Bari - Via E. Orabona 4, 70125 Bari, Italy}, 
\author{F. Longo}
\affiliation{Dipartimento di Fisica, Universit{\`a} degli Studi di Trieste - via A. Valerio 2, 34127 Trieste, Italy}
\affiliation{INFN Sezione di Trieste - via A. Valerio 2, 34127 Trieste, Italy}
\begin{abstract}The Large Area Telescope (LAT) on board the \Fermi Gamma-ray Space Telescope (\Fermi) shows long-lasting high-energy emission in many gamma-ray bursts (GRBs), similar to X-ray afterglows observed by the Neil Gehrels Swift Observatory \citep[\textit{Swift};][]{gehrels2004}. 
Some LAT light curves (LCs) show a late-time flattening reminiscent of X-ray plateaus. We explore the presence of plateaus in LAT temporally extended emission analyzing GRBs from the second \lat GRB Catalog \citep[2FLGC;][]{Ajello2019apj} from 2008 to May 2016 with known redshifts, and check whether they follow closure relations corresponding to 4 distinct astrophysical environments predicted by the external forward shock (ES) model.
We find that three LCs can be fit by the same phenomenological model used to fit X-ray plateaus \citep{Willingale2007} and show tentative evidence for the existence of plateaus in their high-energy extended emission.  The most favorable scenario is a slow cooling regime, whereas the preferred density profile for each GRBs varies from a constant density ISM to a $r^{-2}$ wind environment. We also compare the end time of the plateaus in $\gamma$-rays and X-rays using a statistical comparison with 222 \textit{Swift} GRBs with plateaus and known redshifts from January 2005 to August 2019. Within this comparison, the case of GRB 090510 shows an indication of chromaticity at the end time of the plateau. Finally, we update the 3-D fundamental plane relation among the rest frame end time of the plateau, its correspondent luminosity, and the peak prompt luminosity for 222 GRBs observed by \textit{Swift}. We find that these three LAT GRBs follow this relation. 
\end{abstract}


\section{Introduction}\label{Intro}
GRBs emit in a few seconds the same amount of energy that the Sun will release over its entire lifetime. 
\Swift, launched in November 2004, has observed GRBs within a wide range of redshifts ($z$) from $z=0.085$ to $z = 9.4$ \citep{ kulkarni98, cucchiara11}. More specifically, \Swift, with its on-board instruments - the Burst Alert Telescope; (15 -- 150 keV) \citep[BAT;][]{Barthelmy:05}, the X-ray Telescope (0.3 -- 10 keV) \citep[XRT;][]{2005SSRv..120..165B} and the Ultra-Violet and Optical Telescope (170 -- 650 nm) \citep[UVOT;][]{2005SSRv..120...95R} - provides rapid localization of many GRBs and enables fast multi-wavelength follow-up of the afterglows. The afterglows of GRBs are likely due to an external forward shock (ES), where the relativistic ejecta impacts the external medium \citep{Paczynski1993,Katz+97,meszaros97}. It has already been shown that \Swift GRB LCs have more complex features than a simple power law (PL) \citep{Tagliaferri2005,OBrien2006,Zhang2006,Nousek2006,sakamoto07,Zhang2019}. 
\citet{OBrien2006} and \citet{sakamoto07} showed a flat portion in the X-ray LCs of some GRBs, the so-called “plateau emission”, present right after the decaying phase of the prompt emission. \citet{evans09} found evidence of the plateau in 42\% of X-Ray LCs. The \Swift X-Ray plateaus generally last from hundreds to a few thousands of seconds \citep{Willingale2007}. Physically, this plateau emission has been associated with either the continuous energy injection from the central engine \citep{rees98,dai98,sari2000,zhang2001,Zhang2006,liang2007}, due to the electromagnetic spin down of the so-called magnetars (fast rotating newly born neutron stars) \citep[e.g.,][]{zhang2001,troja07,dallosso2011,rowlinson2013,rowlinson14,rea15,BeniaminiandMochkovitch2017,Toma2007,Stratta2018, Metzger2018} or mass fall-back accretion onto a black hole \citep{Kumar2008,Cannizzo2009,cannizzo2011,Beniamini2017,Metzger2018}.  Furthermore, the plateau has also been associated with reverse shock emission contributions \citep{Uhm2007, genet07}, delayed afterglow deceleration \citep{GranotKumar2006, Beniamini2015}, and off-axis contributions \citep{Beniamini2020reala,Oganesyan2020}. A number of correlations related to the plateau emission have been extensively studied \citep{Dainotti2008, Dainotti2010, Dainotti11a, dainotti11b, dainotti2013a, dainotti15, dainotti2015b, dainotti17, delvecchio16} and applied as cosmological tools \citep{cardone09, cardone10, postnikov14,dainotti2013b}. For reviews on correlations related to the plateau emission and their applications as model discriminators, distance estimators, and as cosmological tools, see  \citet{DainottiDelVecchio2017,dainottibook, Dainotti2018, Dainotti2018b}. Among the plateau correlations, we here mention the existence of a 3--D fundamental plane relation among the luminosity at the end of the plateau, $L_a$, the prompt peak luminosity, $L_{peak}$ and the rest frame time at the end of the plateau, $T_a$ \citep{dainotti16c, dainotti17}.

While \Swift is particularly important for detecting the temporal behaviour of LCs, \Fermi is crucial for detecting the shape of broadband Spectral Energy Distributions (SEDs). The \Fermi Gamma-Ray Burst Monitor \citep[GBM; 8 keV - 40 MeV;][]{Meegan_GBM} has observed more than $2700$ GRBs, conveying new and important information about these sources.
A crucial breakthrough in this field has been the observations of GRBs by the \Fermi Large Area Telescope \citep[LAT; 20 MeV - 300 GeV][]{Atwood:09}. This high-energy emission shows two very interesting features: photons with energy $>100$ MeV peak later \citep{ackermann2013,Ito2013,Ito2014,Ajello2019apj,Omodei+09,Warren2018} and last longer than the sub-MeV photons detected by the GBM. Indeed, the study of three GRBs at energy $>100$ MeV  (080916C, 090510, 090902B), has led to the interpretation that LAT photons are associated with the afterglow rather than the prompt emission, and are generated via synchrotron emission in the ES \citep{Kumar2010,abdo09science,abdo2009apjl,DePasquale2010,Razzaque+10,Meszaros1993,Koveliotou2013,Wang13, meszaros97,waxman:97,Waxman1997b,Omodei+09, Beniamini2015,fraija2020}. In particular, the works of \cite{Koveliotou2013,Wang13, Beniamini2015,fraija2020} have shown that GRBs detected by LAT can be self-consistently modelled using radio, optical, X-ray and sub-GeV LAT observations self-consistently within the ES model.

\citet{Kumar2010} interpreted the observed delay of the $> 100$ MeV emission as related to the deceleration time-scale of the relativistic ejecta. The long lasting duration is interpreted as being due to the PL decay nature of the ES within the context of the standard fireball model. Within this model, the GRB afterglow emission is produced by a population of accelerated electrons with a simple PL, $N(E)\propto {E}^{-p}$ for $p > 2$, where $p$ is the electron spectral index.
They arrived at this conclusion by finding consistency with a closure relation (CR) between the temporal decay index ($\alpha$) of the LCs and the energy spectral index ($\beta$) above $100$ MeV, $\alpha=(3\beta-1)/2$. This relation serves as a rough indication that the observed radiation is being produced in the ES.
An analysis of the LAT LCs by \citet{Omodei2013} shows the presence of breaks, which again can provide possible difficulties for the models mentioned above. These breaks, due to their morphological resemblance with the X-ray afterglow plateaus, may be related to the X-ray plateaus whose existence is well established.

The main question we here answer is whether the LCs of the long-duration LAT emission show similar deviations from a PL (e.g. plateaus) like those seen by XRT at lower energies which detect the prompt emission and the afterglow at BAT hard and XRT soft X-ray wavelengths.
The presence of the above-mentioned breaks in the LAT afterglow data raises the following unexplored and challenging questions which we have investigated here:

\begin{enumerate}
\item How many GRBs observed by LAT show an indication of a flat plateau resembling the X-ray plateaus?
\item Is the emission after the $\gamma$-ray plateaus consistent with the ES emission through testing of their CRs?
\item  Are the $\gamma$-ray and X-ray times at the end of the plateau emissions constant?
\item Do properties from the high-energy emission showing an indication of a plateau follow the 3--D fundamental plane relation?
\end{enumerate}

To answer the first, third, and fourth questions, we analyse the GRB LCs observed by the LAT with a sufficient number of photons to characterize the nature of the deviation from a PL, as well as to determine their LC parameters. To answer the second question, we consider theoretical models that ascribe the X-ray plateau to a continuous, long-lasting energy injection into the ES \citep{zhang01,Zhang2006,Zhang2009,macfadyen01,Zhang+11} or models which suggest a time dependence of the microphysical parameters \citep{BeniaminiandMochkovitch2017} or an off-axis origin of the plateau \citep{Beniamini2020realb, Ryan2020}. We here investigate the existence of the plateau emission among the GRBs observed by the LAT by analysing their LCs as a continuation of our preliminary work presented in the second \lat GRB Catalog (2FLGC; \citet{Ajello2019apj}).  

The paper's structure is as follows: section \S \ref{LAT Sample} shows the data analysis and methodology for \lat and \Swift GRBs, \S \ref{LATresults} the results of the LAT analysis,  \S \ref{the CRs} the $\gamma$-ray CRs and their interpretation, \S \ref{Comparison} the comparison at the end of the plateau emission between \lat and \Swift-XRT, \S\ref{Results from SWIFT and LAT} the results of the 3--D fundamental plane relation including LAT GRBs, and \S \ref{conclusion} a summary of the analysis and results.
\begin{center}
\section{Data analysis and methodology}
\label{LAT Sample}
\end{center}
\subsection{The \Fermi-LAT data analysis} 
We select LAT GRBs observed by \Fermi from August 2008 until August 2016 with observed redshifts. These GRBs are analysed in the 2FLGC (which  includes GRBs from August 2008 until August 2018).
To determine the significance of the detection of sources using maximum likelihood analysis, we define the Test Statistic ($TS$) to be equal to twice the logarithm of the ratio of the maximum likelihood obtained using a model including the GRB over the maximum likelihood value of the null hypothesis, i.e., a model that does not include the GRB. We only include GRBs with a $TS>64$ (19 with redshifts taken from the Greiner web page\footnote{http://www.mpe.mpg.de/~jcg/grbgen.html}) analysed with the new event analysis PASS 8 because this provides a better effective area and energy resolution and consequently allows us to verify the existence of the plateau. Out of these 19 GRBs, we select those that can be fitted with a broken PL in the 2FLGC (13 GRBs). GRB 090323 is removed from our sample because it doesn't have enough flux data points to test our model. Out of the 12 GRBs, 3 have reliable fitting parameters for which the error bars do not exceed the values of the best fit parameters themselves and where the fit converges: GRB 090510, 090902B, and 160509A.

The data preparation procedure is the same as the one adopted in the 2FLGC. We perform a time-resolved analysis fitting the \Fermi-LAT data assuming that the GRB spectrum is described by a simple PL in each time bin, where the number of time bins is the same as the bins presented in the 2FLGC. First, we maximize the likelihood in each time bin obtaining the best value of the spectral index. We derive this spectral index from the standard likelihood analysis with GtBurst. Then, fixing the spectral index to this value, we store the values of the likelihood function for different values of the flux by varying the normalization and integrating the PL in energy and time. There is in some cases spectral evolution from one bin to another, but this evolution is not necessarily a signature of the onset of the plateau. The energy range of this analysis is from 100 MeV to 10 GeV.
We then use the additive property of the log-likelihood to evaluate the sum of the log likelihood for an arbitrary function by simply evaluating the flux in each time bin and obtaining the correspondent value of the log likelihood stored in the previous step. To perform this analysis we use the \texttt{ThreeML} \citep{2015arXiv150203122V,2015arXiv150708343V} package\footnote{Documentation and installation instructions at \url{https://threeml.readthedocs.io/en/latest/index.html}}.
To estimate the best fit parameters we adopt a Bayesian approach: given a data set $D$, a model $M(\theta)$, where $\theta$ is the set of parameters, the posterior probability is given by $P(M(\theta')|D)$. In Bayesian inference, the posterior probability is the product between the likelihood function $L(\theta|D)$, and the prior distribution $P(\theta|\theta')$: $P(M(\theta')|D)$ = $L(M(\theta)|D)P(\theta,\theta')$. In our case, the likelihood is computed by multiplying the value of the likelihood (summing the log likelihood) in each time bin:
\begin{equation}
P(M(\theta')|D)= \prod_{i=1}^{n} L(M(i,\theta)|D)P(\theta,\theta').
\label{eq: likelihhod}
\end{equation}
To model both the prompt and the afterglow emissions, we define our model $M(t,\theta)$ as the sum of two functions $f_{\gamma}(t)$=$f_{1}(t)+f_{2}(t)$ taken from \cite{Willingale2007}, where $f_{1}(t)$ and $f_{2}(t)$ model the prompt and the afterglow emission, respectively. Following the phenomenological model W07, each function $f_{i}$(t) can be written as:
\begin{equation}
f_i(t) = \left \{
\begin{array}{ll}
\displaystyle{K_i \exp{\left ( \alpha_i \left( 1 - \frac{t}{T_i} \right) \right )} \exp{\left (
- \frac{\tau_i}{t} \right )}} & {\rm for} \ \ t < T_i \\
~ & ~ \\
\displaystyle{K_i \left ( \frac{t}{T_i} \right )^{-\alpha_i}
\exp{\left ( - \frac{\tau_i}{t} \right )}} & {\rm for} \ \ t \ge T_i \\
\end{array}
\right .
\label{eq:fc}
\end{equation}

\begin{figure}
    \centering
    \includegraphics{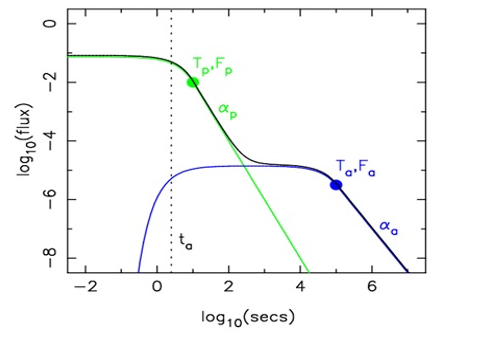}
    \caption{Figure taken from \citet{Willingale2007} illustrating the different parts of the W07 function for clarity.}
    \label{W07}
\end{figure}

\noindent which contains four parameters for the prompt emission ($T_1$,$K_1$,$\alpha_1$,$\tau_1$), and four parameters for the afterglow emission ($T_2$,$K_2$,$\alpha_2$,$\tau_2$) \footnote{$\tau_2=t_a$ in the original notation of the \citet{Willingale2007} model}. The W07 function differs from a broken power law (BPL) fit due to the presence of the exponential in both parts of the equation. The parameter $T_2$ corresponds to the time at the end of the plateau; $K_2$ is the normalization, $\alpha_2$ is the late time PL decay index, and $\tau_2$ is the initial rise timescale. In our sample, all the parameters are free to vary, but we encode prior conditions on $\tau_1$ and $\tau_2$ to allow to fulfill the following conditions: $\tau_1 \leq T_1$, $\tau_2 \leq T_2$, $T_1 \leq T_2$. If these conditions in the fitting are not fulfilled then the fit will give the priors for $\tau_1=T_{LAT,0}$, $\tau_2=T_1$ where $T_{LAT, 0}$ is the estimated LAT emission onset time taken from the 2FLGC. These conditions are similar to the treatment done by \citet{Willingale2007}.
Besides the case of GRB 090902B in which the $\tau_1$ parameter is different from the prior, in all other cases $\tau_1=T_{LAT,0}$. 
We assume a uniform prior distribution for all of the parameters except for $\alpha_1$ and $\alpha_2$, for which we assume a Gaussian distribution centered around the value in the 2FLGC, also shown in Tab.~\ref{FERMITable}, with a $\sigma$ of 2.0.
The Gaussianity hypothesis for $\alpha_1$ and $\alpha_2$ is tested with the values presented in the 2FLGC.
Our analysis uses \texttt{MultiNest} \citep{2008MNRAS.384..449F,2009MNRAS.398.1601F,2019OJAp....2E..10F} to sample the posterior probability and to optimize model parameters.
 
We also check if a simple PL, or the W07 model is the favored fit for our three GRBs. To this end, we first find the minimum value of the two Akaike Criterion information $AIC$ \citep{Akaike2011} statistics pertaining to both the PL, and W07 fit, $AIC_{min}$ = min($AIC_{PL},AIC_{W07}$).  Then, for each model, we calculate the quantity  $B_i=e^{((AIC_{min}-AIC_i)/2)}$, where $B_i$ is the Akaike model weight, and $AIC_i$ and $B_i$ correspond to either the PL, or the W07 fit. Finally, for each model we calculate what is known as the “relative likelihood”, $p_i = B_i/ \Sigma_i (B_i)$. If one of the models has a relative likelihood $p_i>0.95$, we conclude it is significantly favored over the other.
From our analysis, we conclude that the W07 model is favored over the PL models for GRB 090902B and 160509A. Though a PL fit is better for GRB 090510, we continue to keep it in our analysis for the closure relations because it is the only case for which we have a clear plateau emission in X-rays.

Regarding the spectral evolution of the three cases considered, we have calculated the spectral index before, $\beta_{bp}$, during the plateau, $\beta_{dp}$, and after the plateau emission, $\beta_{ap}$. We report the results in Table 1.
In the case of 090510, we see a spectral index consistent with value of 1 before the plateau, steepen by 70\% to become inconsistent with
the value seen before the plateau at more than 2 $\sigma$ during the plateau. In this first case, the spectral index becomes again more shallow after the plateau, reaching a final value again consistent with the starting one. This is not repeated in the case of 090902B, where the spectral indices of the first two phases are consistent with each other at a value of 0.94, but then the spectral index shows an indication of steepening after the plateau to a value of 1.62, albeit with a large
confidence interval of 0.85 at 1 $\sigma$. Finally, 160509A starts off
with a steep spectral index of 2.26 before the plateau, which then becomes much shallower in the final two phases, where it is consistent with a value of 0.5. Thus, the first and third cases show clear signs of spectral index evolution, but not in a consistent manner, while the second GRB is consistent with no evolution. The small sample at hand does not allow the drawing of any definitive conclusion, beyond the evidence for clear spectral index evolution in two out of three cases.

The results of the fits and of the spectral parameters for the 3 \Fermi-LAT LCs are summarized in \S\ref{LATresults}. The results showing that in the cases of GRB 090510 and 160509, when the spectral index change during the plateau leads to another indication that disfavors the energy injection scenario for the plateau.
 
\subsection{The Swift data analysis} 
In addition, we analyse 222 GRBs  with known redshifts detected by \Swift from January 2005 up to July 2019. All of these GRBs have a well-defined plateau in the afterglow phase and a redshift available through \citet{xiao09} and the Greiner web page \footnote{\url{https://www.mpe.mpg.de/jcg/grbgen.html}}, ranging from $z = 0.033$ to $z = 9.4$.  The definition of the plateaus relies on the ability to fit X-ray LCs with the W07 model and obtain results which have reliable error bars through the fitting procedure. The LCs are downloaded from the \Swift web-page repository\footnote{\url{https://www.swift.ac.uk/burst\_analyser}} \citep{evans07, evans09}, and have a signal to noise ratio of 4:1 within the \Swift XRT bandpass $(E_{min}, E_{max})$ = (0.3, 10) keV. We then fit these GRBs with the W07 model, with all of them fulfilling the \citet{avni76} $\chi^2$ prescriptions regarding the determination of the confidence interval (see the XSPEC manual)\footnote{\url{https://heasarc.nasa.gov/xanadu/xspec/manual/XspecSpectralFitting.html}} at the 1$\sigma$ level. 
To obtain the best fit parameters, we use the reduced $\chi^2$ value, which is the $\chi^2$ value divided by the number of degrees of freedom. We note that this is analagous to using a $\chi^2$ or a maximum likelihood test.

\section{Results with LAT data} \label{LATresults}

\begin{figure}
    \centering
     \includegraphics[width=0.7\linewidth]{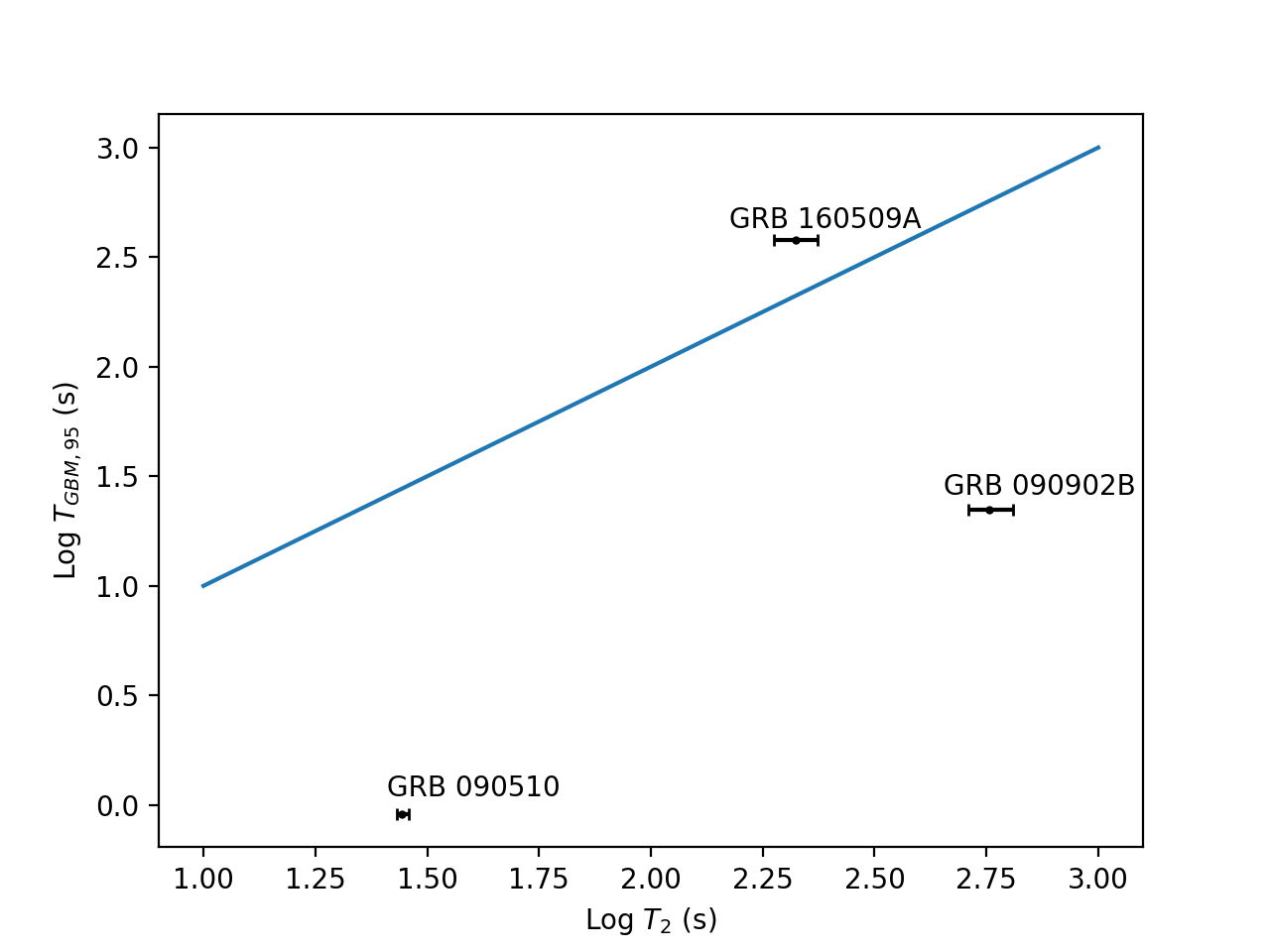}
    \caption{Comparison of $T_2$ derived from the W07 model by fitting the GRBs observed by the LAT versus $T_{GBM,95}$ computed from GBM, with the equality line plotted in blue.}
    \label{fig:my_label}
\end{figure}

Among the \Fermi-LAT GRBs analysed, there are only three cases (090510, 090902B, and 160509A) with known redshift values that have an indication of a plateau according to the fit results.
This analysis allows us to answer the first question regarding the fraction of GRBs with redshift presenting plateaus: $3/19=16\%$. This fraction is $3.7$ times smaller than the fraction of GRBs presenting plateaus in X-rays ($222/373=59\%$) if we consider that \citet{Dainotti2020a} and \citet{Srinivasaragavan2020} performed an analysis spanning from January 2005 until 2019 August. The $\gamma$-ray plateau fraction is also $2.4$ times smaller than the optical plateaus $102/267=38\%$ if we consider the most comprehensive archival analysis of optical plateaus from 1997 to 2016 performed by \citep{Dainotti2020b}. Thus, we can say that the Fermi $\gamma$-ray plateaus are rarer compared to the optical and X-ray plateaus.
Table~\ref{W07parameters} shows that the values of the PL index $\alpha_2$ extracted from the W07 model are all in agreement within 1$\sigma$ with the values of the PL index of the late-time portion of the broken PL model in the 2FLGC. 
In Table~\ref{W07parameters}, we also compare the values of $T_{GBM,95}$ and the values of $T_2$, see Fig. \ref{fig:my_label} where the blue line represents the equality line between the values of $T_{GBM,95}$ and $T_2$. We here stress that $T_{GBM,95}$ is a measure related to the duration of the prompt emission, while $T_2$ is related to the temporal profile of the LC within the W07 phenomenological model. The case of GRB 160509A in particular is peculiar, as its $T_{GBM,95}$ ends significantly later than the end of the plateau $T_2$. For the other  cases, $T_{GBM,95}$ is smaller than $T_2$.

\begin{deluxetable}{r r r r}
\tabletypesize{\footnotesize}
\tablecolumns{4}
\tablewidth{0pt}
\tablecaption{Parameters of the W07 function, along with values of the temporal decay index from the 2FLGC and  $T_{GBM,95}$ start, stop times of the LAT light curves used in the analysis, in some cases $\tau_1=T_{LAT,0}$ and $\tau_2=T_1$, the rest frame isotropic $E_{iso}$ that returns the highest value of the Test Statistics, the redshift and the corresponding distance in Gpc}. Errors are computed using the highest posterior density interval at 68\% confidence level. 
\label{W07parameters}
\tablehead{
\colhead{} &\colhead{090510} &\colhead{090902B} &\colhead{160509A}} 
\startdata
 $K_1 \, (erg \, cm^{-2}\, s^{-1})$ & $(2 \pm 0.5)\times10^{-5}$ & ($5.9 \pm 1.9)\times10^{-6}$ & $(5.4^{+1.5}_{-1.6})\times10^{-7}$ \\
$\alpha_1$& $(2.62^{+0.18}_{-0.17})$ & $2.12^{+0.15}_{-0.16}$ & $3.8 \pm 0.6$ \\
$\tau_1\, (s)$& $T_{LAT,0}$ & $14 \pm 4$ & $T_{LAT,0}$ \\
$T_1\, (s)$& $1.35^{+0.07}_{-0.08}$ & $18.3^{+1.3}_{-1.4}$ & $19.5^{+1.7}_{-1.5}$ \\
$K_2 \, (erg \, cm^{-2}\, s^{-1})$& $(3.97^{+0.27}_{-0.29})\times10^{-8}$ &$(6.8 \pm 1.9)\times10^{-9}$ & $(1.6 \pm 0.4)\times10^{-8}$ \\
$\alpha_2$& $1.41^{+0.31}_{-0.3}$ & $1.5 \pm 0.6$ & $1.64^{+0.31}_{-0.26}$ \\
$\tau_2\, (s)$& $T_1$ & $T_1$ & $T_1$ \\
$T_2\, (s)$& $(27.7^{+1}_{-0.7})$ & $(5.7^{+0.7}_{-0.6})\times10^2$ &$(2.11^{+0.23}_{-0.24})\times10^2$ \\
$F_\gamma(T_2) \, (erg \, cm^{-2}\, s^{-1})$& $(3.78^{+0.72}_{-0.47})\times10^{-8}$ & $(6.76^{+6.9}_{-4.2})\times10^{-9}$  & $(1.46^{+1.35}_{-0.66})\times10^{-8}$ \\
$\beta_{bp}$ & 
$1.00 \pm 0.07$ & $0.92 \pm 0.04$ & $2.26 \pm 0.21$\\
$\beta_{dp}$ &
$1.69 \pm 0.36$ & $0.93 \pm 0.11$ & $0.45 \pm 0.17$\\
$\beta_{ap}$ & 
$1.13 \pm 0.26$ & $1.62 \pm 0.85$ & $0.63 \pm 0.19$\\
\hline
$\alpha_{2FLGC}$& $1.3 \pm 0.2 $ & $1.2 \pm 0.2$ & $1.3 \pm 0.3$ \\
$T_{GBM, \, 95}\, (s)$& 0.91 & 22.14 & 377.86 \\
$T_{LAT, \, 0}\, (s)$& 1.0 & 10 & 15 \\
$T_{LAT, \, 100}\,(s)$& 170.0 & 884 & 5677\\
$E_{iso} \, (erg)$ &$5.1 \times 10^{52}$  & $5.4 \times 10^{53}$  & $5.4 \times 10^{52}$ \\
redshift ($z$) & 0.903 & 1.822& 1.17 \\
distance $(Gpc)$ &  5.86 & 13.94 & 8.07 \\
\hline
\enddata
\end{deluxetable}

\begin{figure}
\begin{center}
\includegraphics[width=0.32\hsize,angle=0]{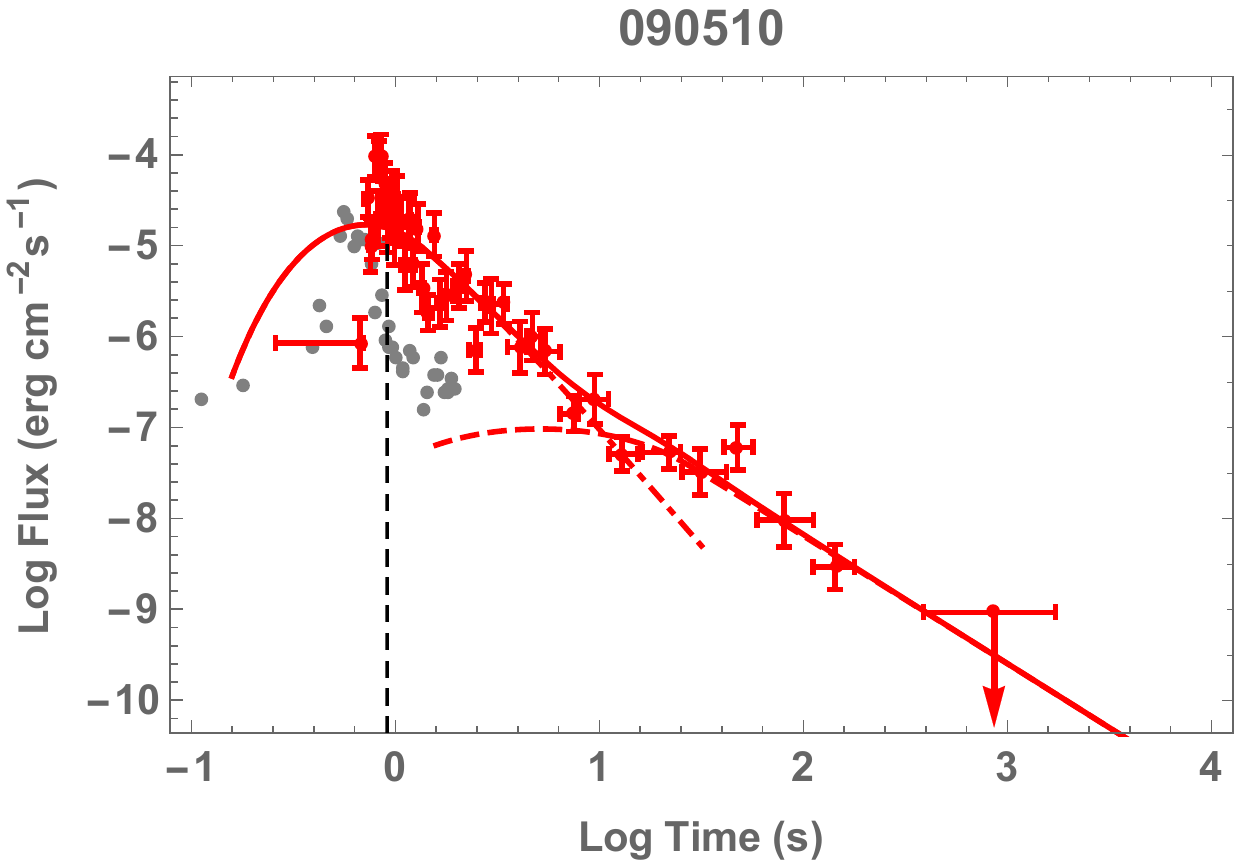}
\includegraphics[width=0.32\hsize,angle=0]{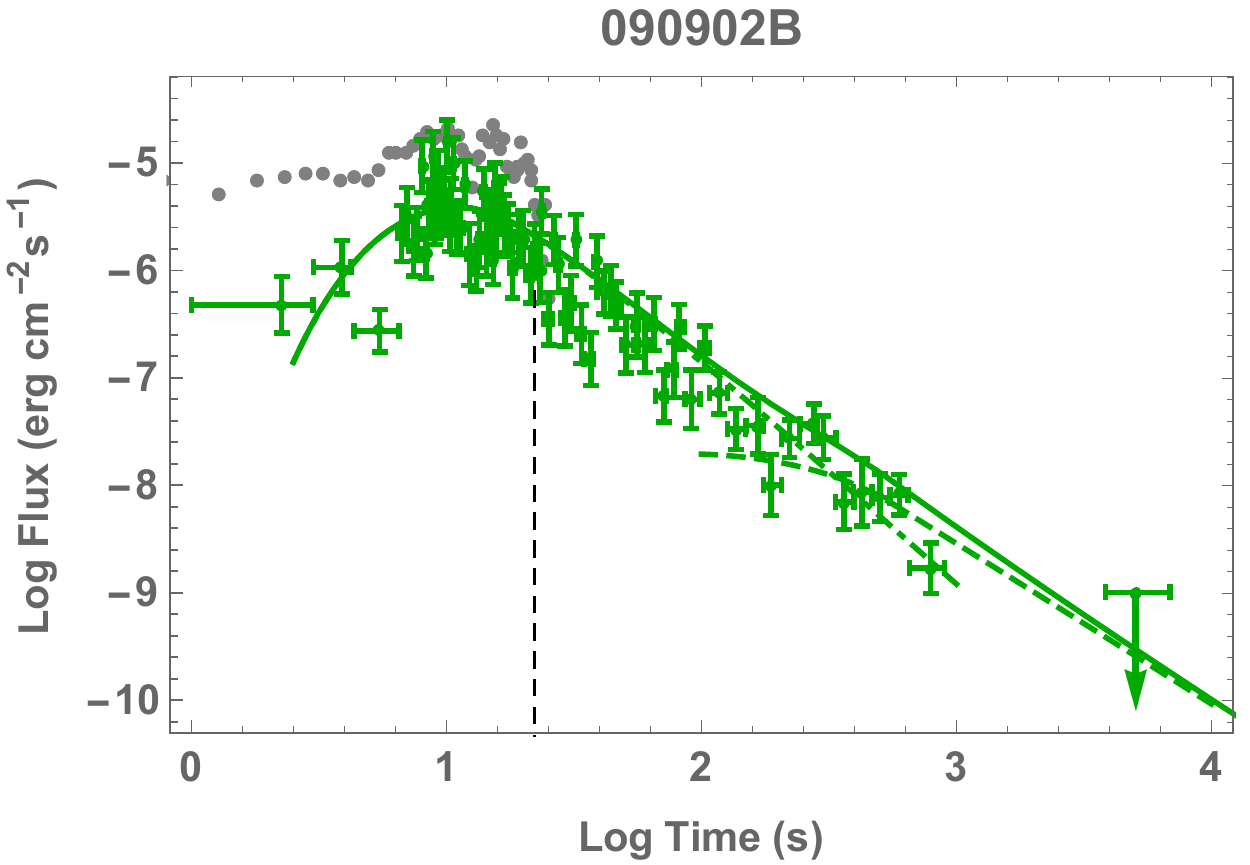}
\includegraphics[width=0.32\hsize,angle=0]{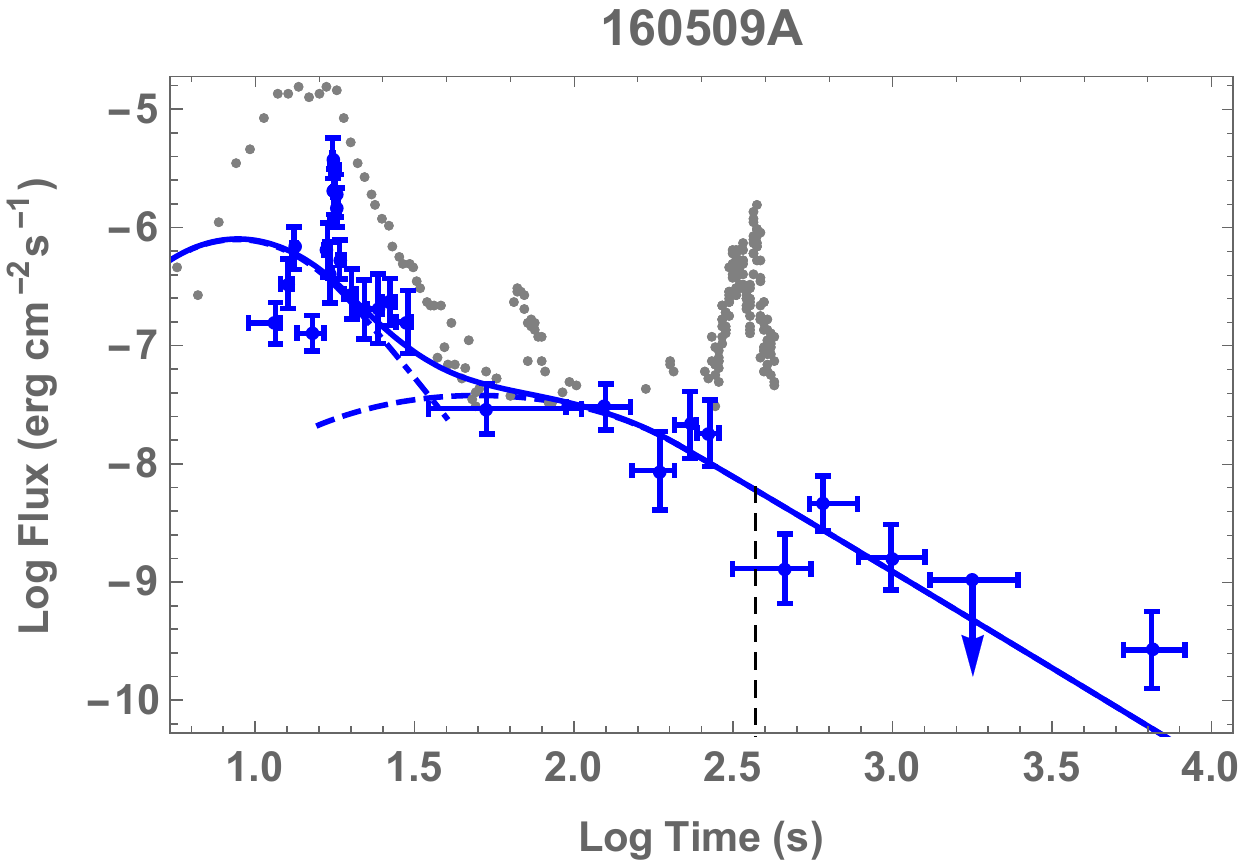}
\caption{Energy flux LCs between 100 MeV and 10 GeV for the 3 LAT GRBs fitted to the W07 model (Eq. \ref{eq:fc}, solid lines). Dot-dashed and dashed lines are the two individual W07 functions $f_{1}(t)$ and $f_{2}(t)$ respectively, the solid line is the sum of the two functions, and the black dashed line indicates the $T_{GBM, \, 95}$. Gray points give GBM LCs for the same GRBs, in the energy range 150 KeV to 30 MeV.}
\label{composite}
\end{center}
\end{figure}

The high-energy flux LCs together with the best fit model for the 3 LAT GRBs are shown in Figure \ref{composite} in colors. The grey data points show the corresponding GBM data.
The results of the fit for the 3  LAT LCs, as well as the parameters of the W07 model and $f_{\gamma}(T_{2}$) are summarized in Table \ref{W07parameters}.
\FloatBarrier

\begin{figure}
\centering
\includegraphics{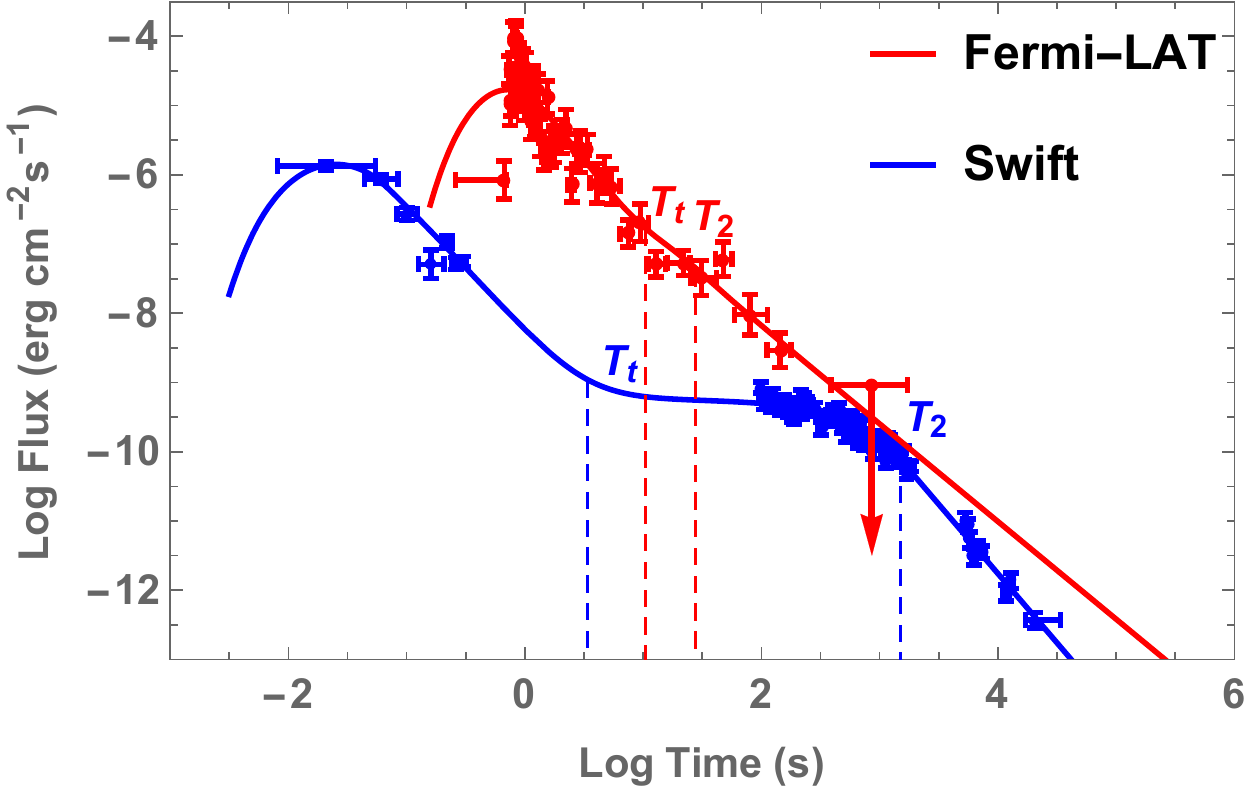}
\caption{Energy flux LC of GRB 090510 using the W07 model in X-rays and $\gamma$-rays, with the corresponding data from \Swift (blue points) and \lat (red points). The vertical dotted lines indicate the corresponding start and end times of the plateau for X-rays (blue) and $\gamma$-rays (red).}
\label{090510plot}
\end{figure}

Among the cases studied, GRB\, 090510 is the only case where the X-ray plateau is also observed in the joint BAT+XRT LCs \citep{DePasquale2010}, see Figure \ref{090510plot}. In this case, the difference between the estimated end times of the plateau phases in X-ray and $\gamma$-rays suggests that the end time of the plateau is not achromatic, which we detail further in Section \ref{Comparison}.

\begin{figure}[h!]
\includegraphics[width=0.5\linewidth]{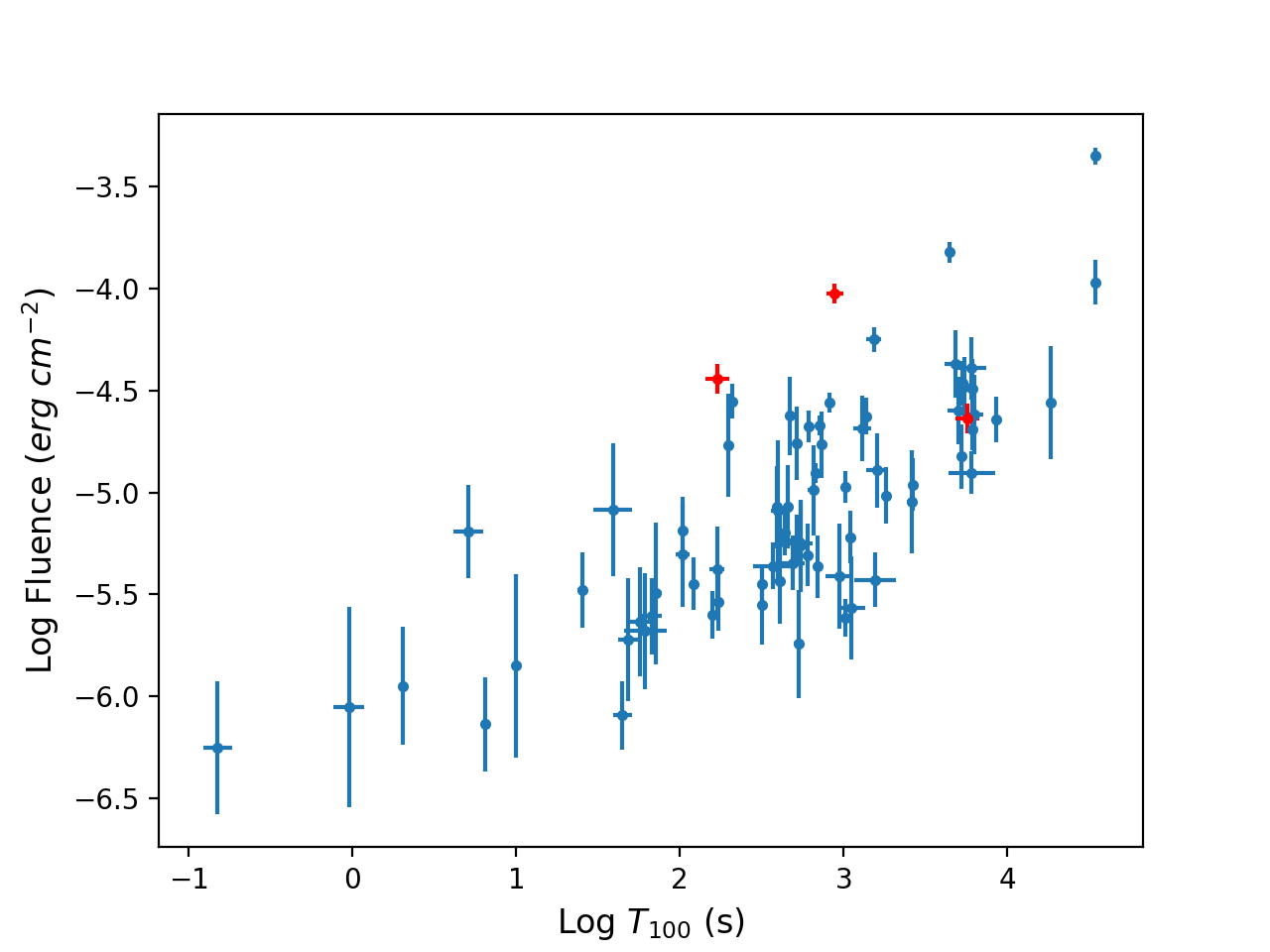}
\includegraphics[width=0.5\linewidth]{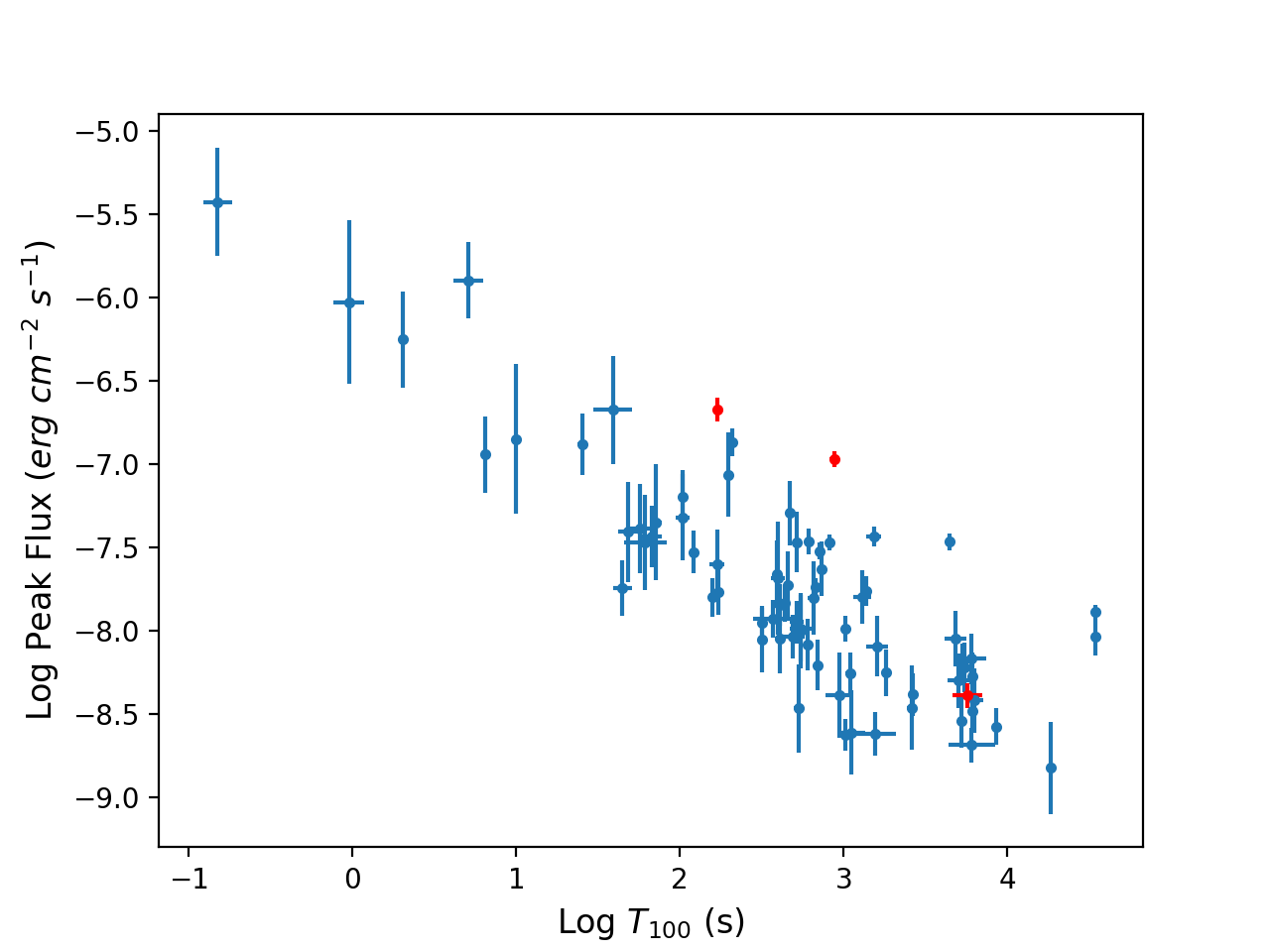}
\caption{Energy fluence and peak energy flux taken from the 2FLGC obtained by the likelihood analysis that returned the highest value of Test Statistics versus $T_{100}$ for 2FLGC LAT GRBs with $TS$ $>$ 64 (blue), along with the 3 GRBs fitted by the W07 model (red).}
\label{fluxfluenceplot}
\end{figure}

In Figure  \ref{fluxfluenceplot} we also show the 100 MeV--10 GeV energy fluence (left panel) and the peak energy flux taken (right panel) from the 2FLGC obtained by the likelihood analysis in the LAT time window as a function of $T_{100}$, which is the time by which 100\% of the high-energy ($>$ 100 MeV) photons associated with a GRB are detected (from the 2FLGC).
As expected, the 3 GRBs are all placed at the high end of the $T_{100}$ distribution, with fluence and flux values higher than $2.3 \times 10^{-5}$ erg cm$^{-2}$ and $4.08 \times 10^{-9}$ erg cm$^{-2}$ s$^{-1}$, respectively. 
It is likely that for GRBs with lower fluences/fluxes, the sensitivity of the detector does not allow the plateau to be detected. On the other hand, a few bright GRBs in the 2FLGC do not show any significant deviations from a simple PL fit, suggesting that high-energy plateaus are not a universal characteristic of these GRBs.



\FloatBarrier

\section{Test of CRs}\label{the CRs}

There are multiple proposed CRs associated with the ES model, although the uncertainty of the observations do not allow us to differentiate between all relations. We here discuss nine of these CRs corresponding to 4 distinct astrophysical environments, where we do not consider post-jet break relations or relations due to energy injection into the ES.

According to \citet{Zhang2006} and \citet{Nousek2006}, X-ray afterglow LCs have four emission phases. Following \citet{Racusin+09}, we consider LCs divided into several segments which follow the forms as described by \citet{Zhang2006} and \citet{Nousek2006}. According to \citet{Racusin+09}, these 4 segments are: I) the initial steep decay often attributed to the high-latitude emission or the curvature effect \citep{Kumar+00,Qin2004,liang06,zhang07b}; II: the plateau whose origin and features have already been discussed in the introduction; III: the normal decay due to the deceleration of an adiabatic fireball \citep{m02,Zhang2006}; IV: the post-jet break phase \citep{Rhoads99,sari99,m02,piran05}. Flares are seen in around 1/3 of all Swift GRB X-ray afterglows during any phase (I-IV), and may occur due to sporadic emission from the central engine \citep{Burrows05,Zhang2006,Falcone2007}.

The CRs are relations between the temporal and spectral PL indices ($\alpha$ and $\beta$) that probe the physical details of the ES fireball model, assuming that synchrotron radiation is the dominant mechanism in the afterglow. These correlations vary according to the emission processes occurring in the part of the afterglow LC we focus on, the surrounding environment, the electron spectral distribution, cooling regime, and jet geometry \citep{MeszarosRees:94,meszaros97,Sari+98,Chevalier2000,Dai&Cheng2001,m02,Zhang2004,piran04}.
The electron spectral index, $p$, is typically $ > 2$. However, a value of $p< 2$ can explain
observations of shallow temporal decays \citep{Panaitescu2001,Bhatt2001}. Therefore, we include all alternatives similarly to \citet{Racusin+09}. 

In our analysis, we only consider phase III of the normal decay phase for LCs that present a plateau emission since we are interested in investigating this particular region. We do not consider the post-jet break phase (IV) because the jet break occurs at very late times compared to the time interval of the high-energy emission. 
We then tested the CRs \citep[given by][]{Racusin+09} for the 3 LAT GRBs, in order to answer question $2$ from \S \ref{Intro} and check whether the normal decay phase after the $\gamma$-ray plateaus obeys the ES emission model. The CRs tested are derived from either a constant density interstellar medium (ISM, $n = constant$) or a wind medium ($n \propto{r^{-2}}$) assumptions, both with no energy injection. Each CR is also characterized by its electron spectral index value (whether $1<p<2$ or $p>2$) and by whether the cooling is in the slow or fast regime. Table~\ref{CR} summarizes the CRs  tested in this paper, with the corresponding physical scenario. Note that for each GRB, we test only the relations corresponding to the respective $p$ value of each GRB, derived through the relation $\beta(p)$ detailed in Table \ref{CR}.

\begin{figure}
\centering
 \includegraphics[width=1.0\columnwidth]{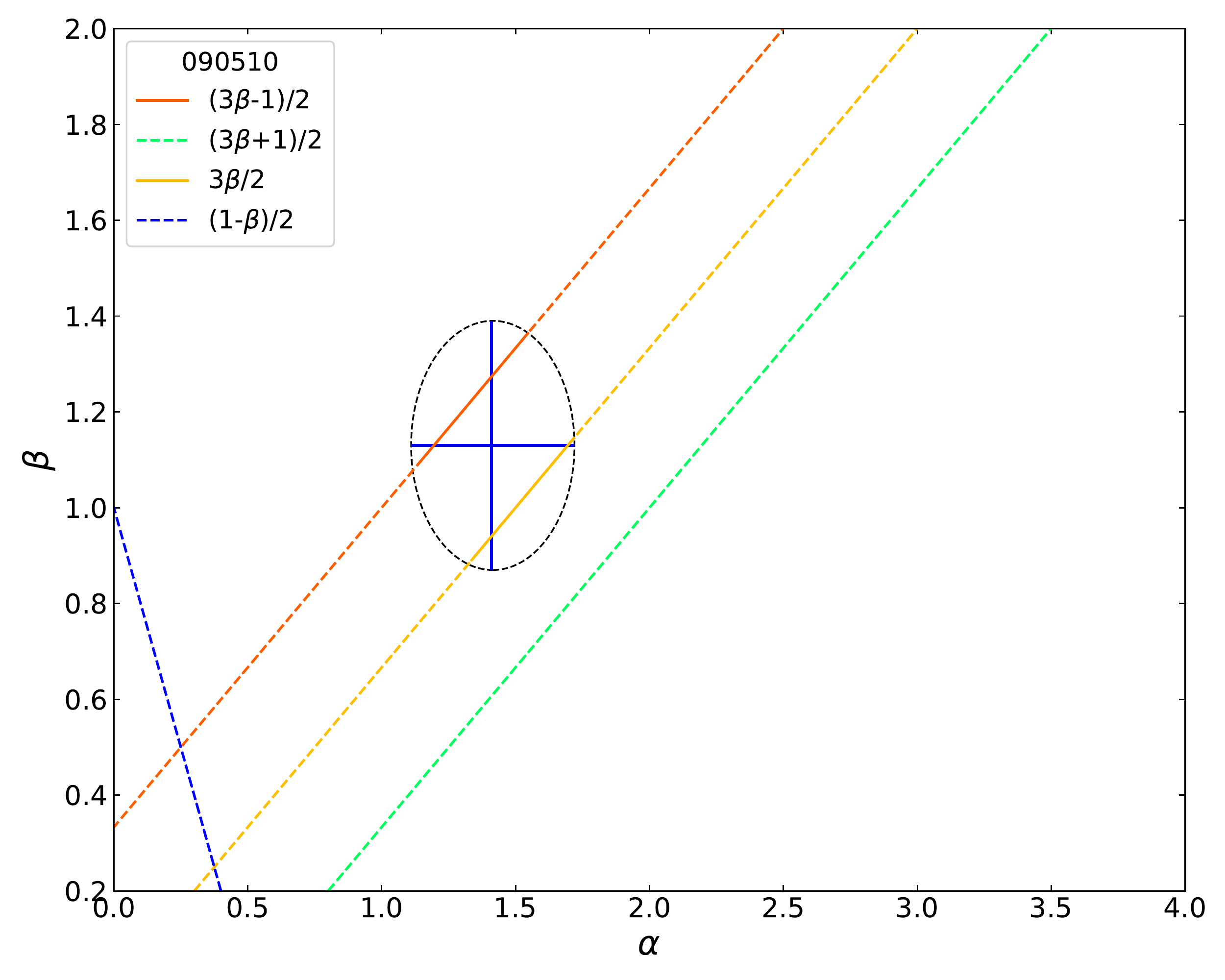}
 \caption{CRs for GRB\,090510. Each line corresponds to a particular CR and is colored from red to light blue depending on the distance of the measured $\alpha$, $\beta$ parameters from the actual equality line of the CR. The ellipse indicates the error of $\alpha$ and $\beta$ and it is used to calculate whether a particular CR is fulfilled.}
  \label{plot:closure090510}
\end{figure}

\begin{figure}
\centering
 \includegraphics[width=1.0\columnwidth]{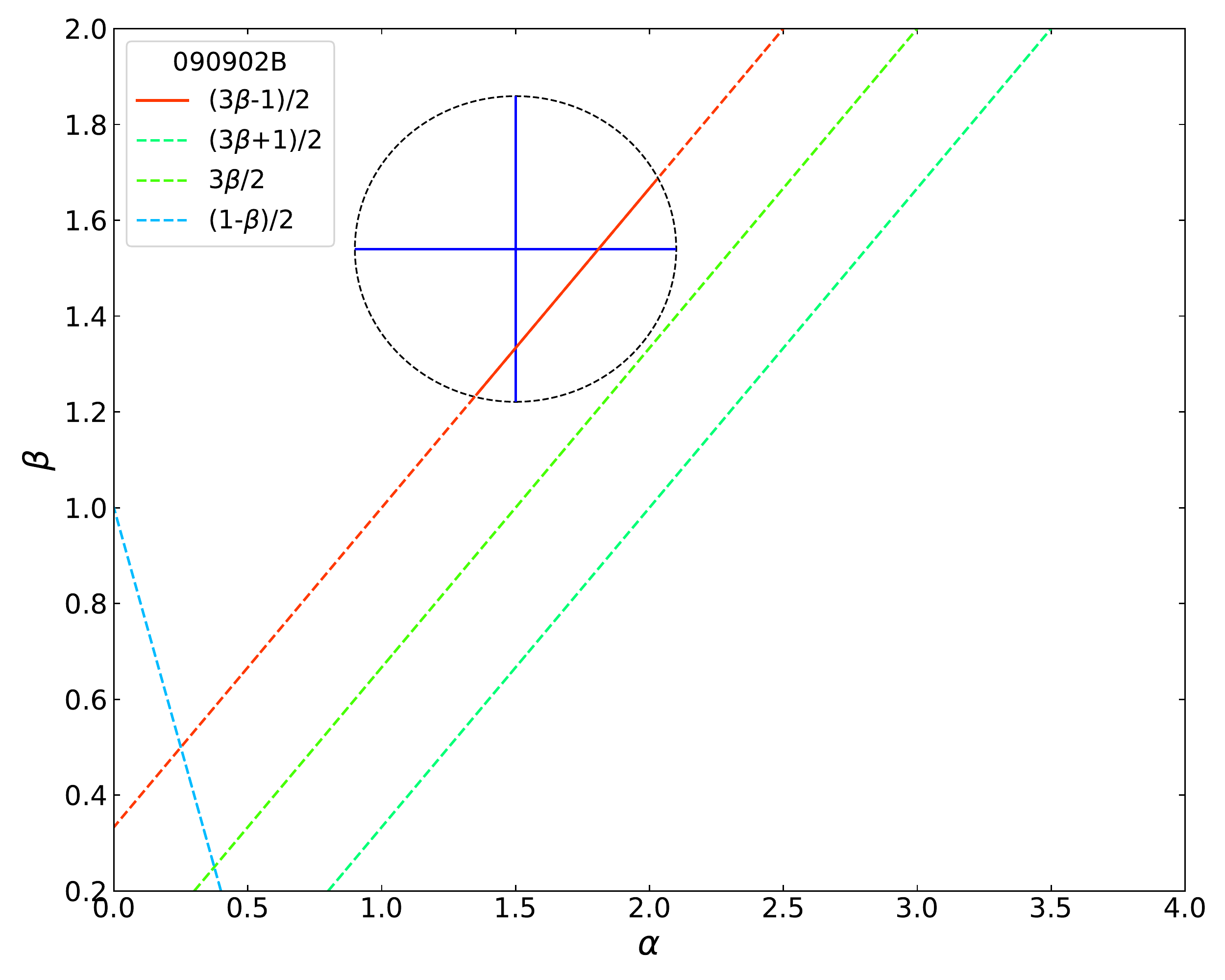}
 \caption{CRs for GRB\,090902B. See caption of Figure ~\ref{plot:closure090510} for details.}
  \label{plot:closure090902B}
\end{figure}

\begin{figure}
\centering
 \includegraphics[width=1.0\columnwidth]{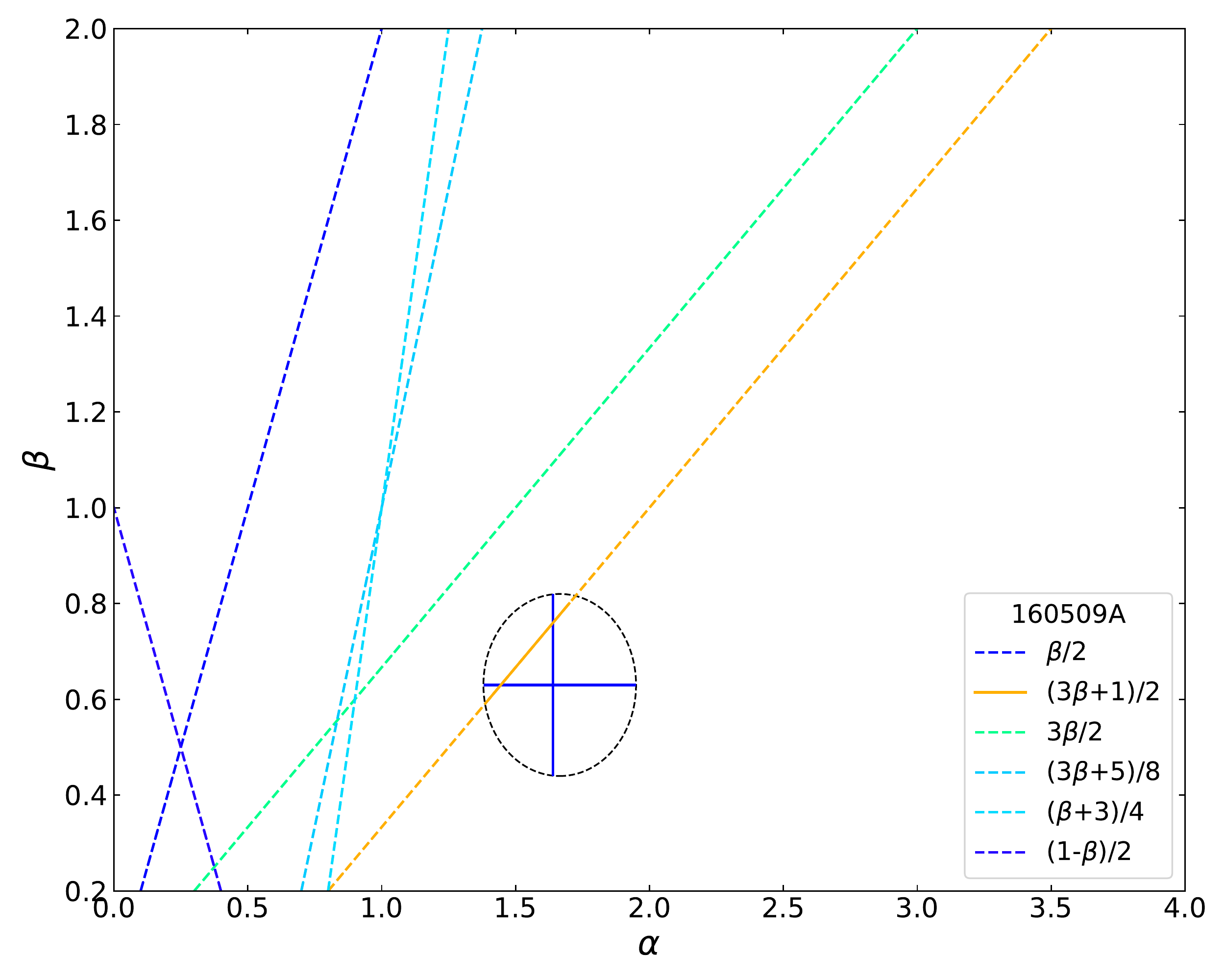}
 \caption{CRs for GRB\,160509A. See caption of Figure \ref{plot:closure090510} for details.}
  \label{plot:closure160509A}
\end{figure}

The outcomes of this analysis are shown from Figure  \ref{plot:closure090510} to Figure \ref{plot:closure160509A}.  The figures show each GRB in the ($\alpha$, $\beta$) plane, with all the CRs represented as lines.
Table \ref{FERMITable} summarizes all of our results. Note that we propagate the errors of $\alpha$ and CR($\beta$) by sampling the posterior distribution and testing whether or not a given CR is satisfied using the highest posterior density interval at a 68\% confidence level. For doing so, we assume an elliptical shaped error in the $\alpha$, CR($\beta$) plane and determine if the assumed CR crosses the ellipse. Graphically, each CR is colored depending on the distance from the GRB. In Table \ref{FERMITable}, we detail the closure relationship values and at which $\sigma$ level each GRB fulfills these relations. Note that the value of the decay index $\alpha$ corresponds to the value of $\alpha_2$ of our model, while the value of the spectral index $\beta$ needs to be calculated using GtBurst from the end time of the plateau $(T_2)$ to the end time of the LC (where the last bin is detected with a signal above 3 $\sigma$, see the 2FLGC for details).

Some interesting features appear when analysing the results of the CRs. There are no CRs that are fulfilled for all three GRBs. Note that when we mention the phrase ``fulfilled" hereafter, we are referring to a fulfillment within a 1$\sigma$ level. The relations  that are not fulfilled by any GRBs that are still in the GRBs' respective $p$ ranges are  $\alpha=(1-\beta)/2$, which is not fulfilled for any of the GRBs, and $\alpha=\beta/2$ and $\alpha=(\beta+3)/4$, which are not fulfilled by 160509A. This implies that a fast cooling environment is not favored for our set of GRBs. The only relation which is fulfilled by multiple GRBs is $\alpha=(3\beta-1)/2$, which is fulfilled by 090902B and 090510. This relation  implies a slow cooling environment, consistently lining up with the previous comment that a slow cooling environment seems to be preferred for our set of GRBs.

In the case of GRB\,090510 (Figure \ref{plot:closure090510}), 2 closure relations are fulfilled:  $\alpha=(3\beta-1)/2$ and $\alpha=3\beta/2$. Through this, we infer that either an ISM or wind environment with slow cooling is preferred. However, X-ray afterglow data shows that the medium in the vicinity of the burst must be of constant ISM \citep{Kumar2010} with either slow or fast cooling. Therefore, we can conclude that the most likely scenario for this GRB is a constant ISM environment with slow cooling. For the case of GRB\,090902B (Figure \ref{plot:closure090902B}) 1 CR is fulfilled: $\alpha=(3\beta-1)/2$. We see again that this relation corresponds to either an ISM or wind environment with slow cooling. However, \citet{Ajello2018} came to the conclusion that FERMI-LAT is biased towards detecting GRBs that occur in lower wind density environments. This is due to the fact that the synchrotron cooling-break for GRBs in a wind environment occurs at high energies, and does not deviate down to lower energies like it does for GRBs in a constant density ISM environment. This makes the afterglow spectrum in $\gamma$-rays and hard X-rays last for longer time periods, making it more likely that LAT will detect these GRBs \citep{Ajello2018}. Therefore, though our analysis of the closure relations points towards either an ISM or wind environment with slow cooling for GRB 090902B, we conclude with the help of \citet{Ajello2018} that a wind environment with slow cooling is preferred. For GRB\,160509A (Figure \ref{plot:closure160509A}), 1 CR is fulfilled: $\alpha=(3\beta+1)/2$. This relation corresponds to a wind environment with slow cooling, making it the most likely scenario of the ones tested \citep{Warren2021}:

\begin{equation}
\begin{aligned}
\epsilon_{B} \lesssim (\frac{h\nu_{max}}{660 \times n_{ISM,0}^{-0.18} \times E_{iso,53}^{0.03} \times (T_{obs}/100)^{0.008}})^{-5.49},\\
\end{aligned}
\end{equation}

where $E_{iso,53}$ is the isotropic energy divided by $10^{53}\,  \mathrm{ergs}$ $n_{ISM,0}$ is the particle density in the ISM, which we assume to be $1$ particle per $\mathrm{cm}^3$, and $h\nu_{max}$ is the maximum frequency which we assume to be 1 $\mathrm{GeV}$. With this requirement we obtain $\epsilon_b=0.095, 0.11, 0.11$ for 090510, 090902B and 160509B, respectively.

\subsection{Comparison with the analysis of Tak et al. (2019) and Kumar \& Barniol (2010)}

When we compare our results with those of \citet{Tak2019} we find a few differences. However, these differences are well justified, since \citet{Tak2019} used a simple PL fit, as opposed to our 4-parameter W07 model for the afterglow. As a result, our $\alpha$ temporal indices are different, leading to differences in the outcomes of the CRs. 
To compare our results with the ones obtained by \citet{Tak2019}, we focus on the CRs discussed in both analyses.

For GRB\,090510, the most favorable model in the analysis of \citet{Tak2019} relies on the CR $\alpha=3\beta/2$, which is fulfilled within our analysis, showing that we have reached consistent results for this particular GRB inferring a constant ISM environment with slow cooling.

For GRB\,090902B, \citet{Tak2019} finds no favorable model with their analysis, while 1 of our common CRs, $\alpha=(3\beta-1)/2$, is fulfilled in our analysis. In this particular case, the presence of the plateau may possibly lead to the discrepancy between our results and \citet{Tak2019}. Also suggestive of the above considerations, according to \citet{Tak2019}, an energy injection scenario for this GRB may be required.

For GRB\,160509A, the most favorable model according to the analysis of \citet{Tak2019} is $\alpha=(3\beta-1)/2$, which is not in the suitable $p$ range given our value of $\beta$. For this particular GRB, \citet{Tak2019} also suggests an alternative model, $\alpha=3\beta/2$, which is also not fulfilled by our analysis at a $1\sigma$ level, though it is fulfilled at a 2$\sigma$ level. So, if we assume the absence of a plateau and a simple PL fit, the most probable scenario for this GRB is either an ISM or wind environment with fast cooling.

When we compare our results with the ones presented in \citet{Kumar2010}, for GRB\,090510, \citet{Kumar2010} found that the CRs $\alpha=3\beta/2$ and $\alpha=(3\beta+5)/8$ are both fulfilled. In our case, $\alpha=3\beta/2$ is fulfilled, while $\alpha=(3\beta+5)/8$ is not in the suitable $p$ range given our value of $\beta$. 
The differences for GRB 090510 can be explained due to the prompt sub-MeV emission not being considered in \citet{Kumar2010}, so the first few points of the LC were not included in their fit. In addition, we use the PASS 8 \citep{Atwood13} event class for our analysis (which did not exist in 2010,
when \citet{Kumar2010} was written), and which improves the precision of our fitting.
We also do not have compatible results for GRB\,090902B, since the $\alpha=3\beta/2$ CR is not fulfilled by our analysis, while it is in theirs. 


\subsection{Comparison with the analysis of Maxham et al. (2011)}

When comparing our analysis with that of \citet{Maxham2011}, there are two GRBs in common: GRB 090510 and GRB 090902B.
\citet{Maxham2011} calculated energy accumulation in the external shock by assuming a constant radiative efficiency. By solving for the early evolution of both an adiabatic and a radiative blast wave, they compute the high-energy emission in the \lat band and compare it with the observed one for the above mentioned GRBs. 
The late time \lat LCs after $T_{90}$ can be fitted by their model. However, due to continuous energy injection into the blast wave during the prompt emission phase, the early ES emission cannot account for the observed GeV flux level. They reached the conclusion that the high-energy emission during the prompt phase (before $T_{90}$) may derive from two components: a rising ES component and a dominant component of an internal origin. 
According to \citet{Maxham2011} a simple broken PL LC is expected from the blast wave evolution of an instantaneously injected fireball with a given initial energy. Such an approximation holds if the time scale at play is much longer than $T_{90}$. However, during the time in which the central engine is still active, they do not predict a simple LC evolution, since the energy output from the central engine is continuously injected into the blast wave.
To properly follow this reasoning and make a meaningful comparison of our results and the interpretation of \citet{Maxham2011} we refer to Table 1, where we quote the values of $T_{GBM, \, 95}$ (we use $T_{GBM, \, 95}$ instead of $T_{90}$ since it takes into account a larger percentage of the energy emitted in the prompt emission)  and $T_2$ with its error bars.
In our analysis, GRB 090510 and GRB 090902B show that for the time at the end of the plateau emission $T_2 \gg T_{95}$. 

The case of GRB 160509A, where $T_{95} \gg T_2$, is different. This may mean that a different mechanism from the external shock must be conceived. For example, one possibility is energy injection. 
In principle, a plausible explanation for the energy injection could be magnetar emission. It is known that the validity of the magnetar scenario relies on a maximum energy of $3 \times 10^{52}$  erg for a “standard” neutron star, with a mass of $1.4 \, M_{\odot}$ and a radius of 12 km, with a minimum spin period of 1 ms \citep{Bucciantini2007, Bucciantini2009, Duffell2015, Lattimer2016}. However, in more massive ($2.1 \, M_\odot$) and compact ($R \approx 10$ km) neutron stars, the maximum spin energy can reach up to $10^{53}$ erg \citep{Dall2018}. 
We also acknowledge that there is an ongoing discussion about the limiting energy for powering a GRB from a magnetar. For example, \citet{BeniaminiMetzer2017} pointed out that such a value of $10^{53}$ erg cannot be reached, since only a fraction of the energy released by a magnetar can have an energy per baryon ratio that is able to account for the bulk Lorentz factor of the ejecta that is required by the problem of compactness.

GRB 160509A has $E_{iso_{LAT}}=1.0 \times 10^{53}$ erg (from 2FLGC), so the required values for the spin period may still be physically possible (though very small), of around 0.51 ms \citep{Dall2018}. Thus, it would be extremely interesting to study this GRB within the magnetar model in a future study.
On the other hand, the fact that the LAT CRs after the end of the plateaus are consistent with the external shock scenario, in general means that an extra energy injection is not needed. Indeed, as we have already mentioned in the introduction we can contemplate models which require a temporal dependence of the microphysical parameters or an off-axis origin of the plateau emission. Both these scenarios predict the exact same closure relations in phase III of the ES afterglow. Several of the other potential mechanisms for plateaus mentioned in the introduction would also satisfy this observation.

\subsection{Interpretation of the results obtained with the CRs}

From the test of a set of nine CRs, we found that the normal decay phase after the plateau is consistent with the ES scenario. This result supports that the late-time high-energy emission of \lat GRBs originates from the ES scenario, and opens a new possibility for understanding the high-energy emission in a wider time scale\footnote{We here would like to mention as a caveat that it is sometimes hard to assess the validity of the closure relations due to possible problems encountered in reliably fitting the LCs, see \citet{Gulli2006} for details.}.
In our analysis, we consider linear particle acceleration, although the non-linear particle acceleration scenario \citep{2017ApJ...835..248W} cannot be ruled out. Also, other mechanisms could explain the high-energy emission after the plateau phase.

We can interpret the results of the CRs
within the four following main scenarios:
\begin{enumerate}
\item The three LAT GRBs showing an indication of a plateau could all be fit by at least one of the CRs derived from the ES model. Hence, the standard fireball model and the ES scenario are a suitable explanation for this small set of GRBs, for a subset of the CRs tested.

\item CRs are a quick check to assess the reliability of the ES scenario, thus it is possible that the ES is still the most viable explanation. However, it is also possible that the ES formulation is lacking some details, especially at high energies and in the presence of a plateau phase.
In this context, the fully radiative solution proposed by \citet{Maxham2011} can account for the observed LCs of two GRBs in the sample.

\item In the afterglow, non-linear particle acceleration can occur. \citet{2017ApJ...835..248W} studied the time evolution of afterglow LCs by taking into account effects of non-linear particle acceleration for the first time. They found that temporal and spectral evolution is much different from the formulation of the linear particle acceleration afterglow model mentioned above.
Also, \citet{2017ApJ...835..248W} showed that very high-energy $\gamma$-rays can be produced by SSC, especially at the early phase of the afterglow. 

\item Given that these three aforementioned  scenarios can coexist, the ES scenario is a possibility for our set of GRBs. It follows that for GRBs with a more complex morphology or spectral features, the CRs examined may be too simplistic. On the other hand, more complex scenarios such as the one mentioned above can more accurately model plateaus in high-energy GRB LCs.

\end{enumerate}
Thus, the most plausible interpretation is that these four scenarios may separately occur for a set of GRBs. The ES model can still be a good explanation of the high-energy LCs presenting a plateau emission. However, we must remain open to exploring new possibilities which allow us to verify if the cases that do not follow the ES model are pinpointing the presence of non-linear particle acceleration, an energy injection mechanism such as the one obtained with a magnetar, models which rely on the variation of the microphysical parameters, or models that require the plateau emission being generated off-axis. More and higher quality data can probably help us to continue shedding light on these results in the near future.  
\FloatBarrier
\begin{deluxetable}{cccccc}[h!]
\tablecolumns{4} 
\tablewidth{0pc} 
\captionsetup{justification=centering}
\caption{Closure relations (Part of the table is taken from \citet{Racusin+09})}
\label{CR}
\tabletypesize{\footnotesize}
\tablehead{ 
\multicolumn{4}{c}{No Energy Injection} \\ 
\cline{1-4}
\colhead{$\nu$ range} & $\beta(p)$ & \colhead{$\alpha(\beta)$} & \colhead{$\alpha(\beta)$} & \\
& & \colhead{$(p > 2)$} & \colhead{$(1<p<2)$}}
\startdata 
\multicolumn{4}{c}{ISM, Slow Cooling} \\[0cm]
\cline{1-4} $\nu_m < \nu < \nu_c$ & $\frac{p-1}{2}$ &
$\alpha=\frac{3\beta}{2}$  & $\alpha=\frac{3(2\beta+3)}{16}$\\ 
$\nu > \nu_c$& $\frac{p}{2}$ & $\alpha=\frac{3\beta-1}{2}$ &  $\alpha=\frac{3\beta+5}{8}$
\\
\cline{1-4}
\multicolumn{4}{c}{ISM, Fast Cooling} \\[0cm]
\cline{1-4}
$\nu_c < \nu < \nu_m$ & $\frac{1}{2}$ &
 $\alpha=\frac{\beta}{2}$ &
 $\alpha=\frac{\beta}{2}$ &   \\
$\nu > \nu_m$ & $\frac{p}{2}$&
 $\alpha=\frac{3\beta-1}{2}$ &
 $\alpha=\frac{3\beta+5}{8}$ &  \\
\cline{1-4}
\multicolumn{4}{c}{Wind, Slow Cooling} \\[0cm]
\cline{1-4}
$\nu_m < \nu < \nu_c$ & $\frac{p-1}{2}$ &
 $\alpha=\frac{3\beta+1}{2}$ &
 $\alpha=\frac{2\beta+9}{8}$ \\
$\nu > \nu_c$ & $\frac{p}{2}$ &
 $\alpha=\frac{3\beta-1}{2}$ &  $\alpha=\frac{\beta+3}{4}$
\\
\cline{1-4}
\multicolumn{4}{c}{Wind, Fast Cooling} \\[0cm]
\cline{1-4}
$\nu_c < \nu < \nu_m$ & $\frac{1}{2}$ &
$\alpha=\frac{1-\beta}{2}$ &
 $\alpha=\frac{1-\beta}{2}$ \\
$\nu > \nu_m$ & $\frac{p}{2}$ &
 $\alpha=\frac{3\beta-1}{2}$ &  $\alpha=\frac{\beta+3}{4}$\\
\cline{1-4}
\enddata
\end{deluxetable}


\FloatBarrier
\begin{deluxetable}{lrrrrr}
\tablecolumns{6}
\tablewidth{0pt}
\tablecaption{Temporal ($\alpha$) and spectral ($\beta$) indices along with the CRs for each of the LAT GRBs. For each GRB in our analysis, we show the $\sigma$ level for which the relations are satisfied. See text for details.} \label{FERMITable}
\tablehead{
\colhead{} 
&\colhead{GRB 090510} &\colhead{GRB 090902B}&\colhead{GRB 160509A}}
\startdata
$\alpha$ & $1.5 \pm 0.6$ & $1.41^{+0.31}_{-0.30}$ &  $1.64^{+0.31}_{-0.26}$\\ 
$\beta$ & $1.13 \pm 0.26$ & $1.54 \pm 0.319$ &  $0.63 \pm 0.19$ \\ 
$\alpha=(3\beta-1)/2$ & $1.20 \pm 0.39$ $< 1\sigma$ & $1.81 \pm 0.48$ $< 1\sigma$ &  -\\
 $\alpha=\beta/2$ & - & - & $0.315 \pm 0.10$ $<5\sigma$\\
 $\alpha=(3\beta+1)/2$ & $2.20 \pm 0.39$ $<2\sigma$& $2.81 \pm 0.48$ $< 2\sigma$ & $1.45 \pm 0.29$ $<1\sigma$\\
$\alpha= 3(2\beta+3)/16$ & -& - & -\\
 $\alpha=3\beta/2$ & $1.70 \pm 0.39$ $< 1\sigma$& $2.31 \pm 0.48$ $< 2\sigma$ &  $0.95 \pm 0.29$ $<2\sigma$\\
 $\alpha=(3\beta+5)/8$ & - & - & $0.86 \pm 0.07$ $<3\sigma$\\
 $\alpha=(2\beta+9)/8$ & - & - & -\\
 $\alpha=(\beta+3)/4$ & - & -& $0.91 \pm 0.05$ $<3\sigma$\\
 $\alpha=(1-\beta)/2$ & $-0.07 \pm 0.13$ $<4\sigma$ & $-0.27 \pm 0.16$ $< 4\sigma$ &  $0.19 \pm 0.10$ $<5\sigma$ \\
\hline
\enddata
\end{deluxetable}

\begin{figure}[h!]\begin{center}
\includegraphics[width=.75\linewidth]{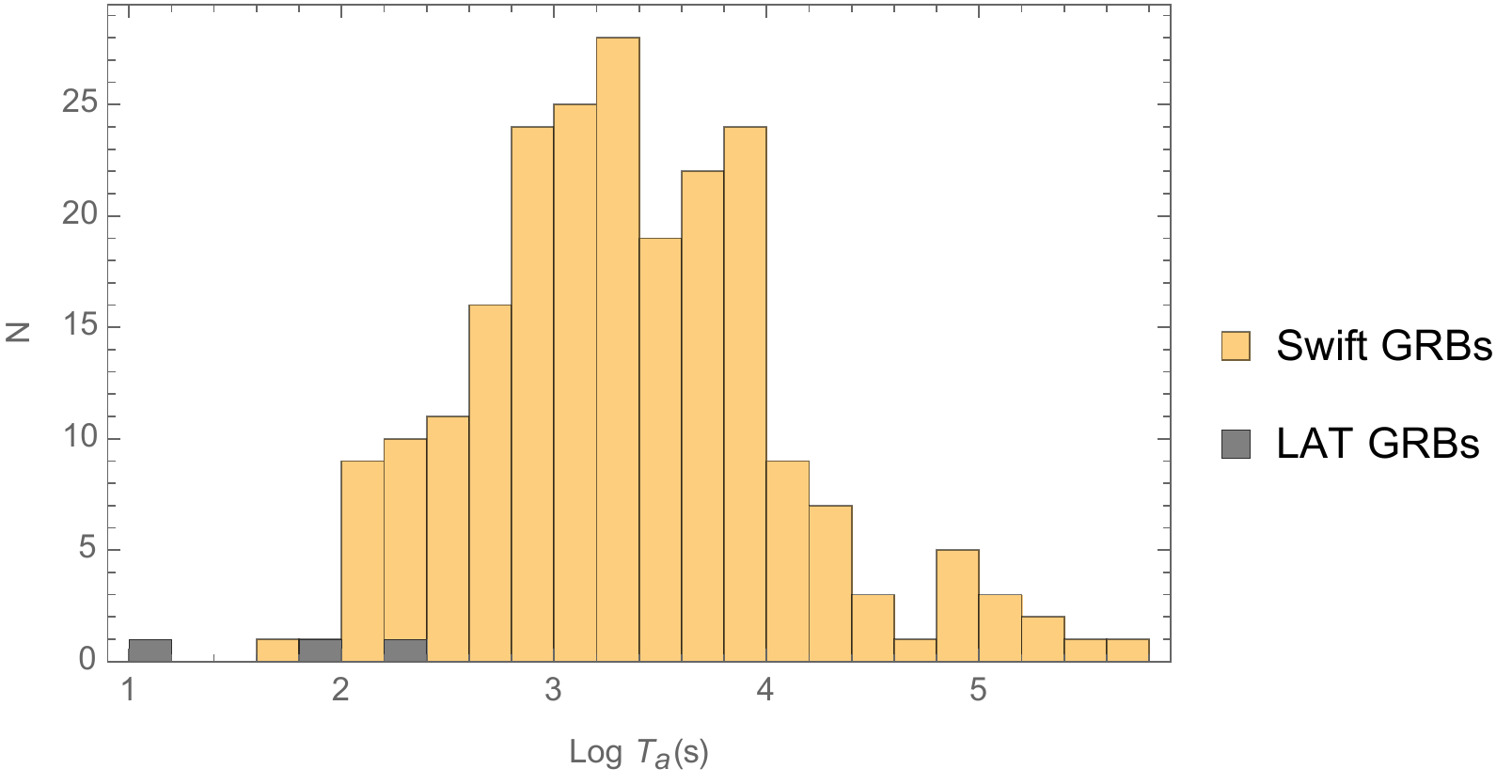}
\includegraphics[width=.75\linewidth]{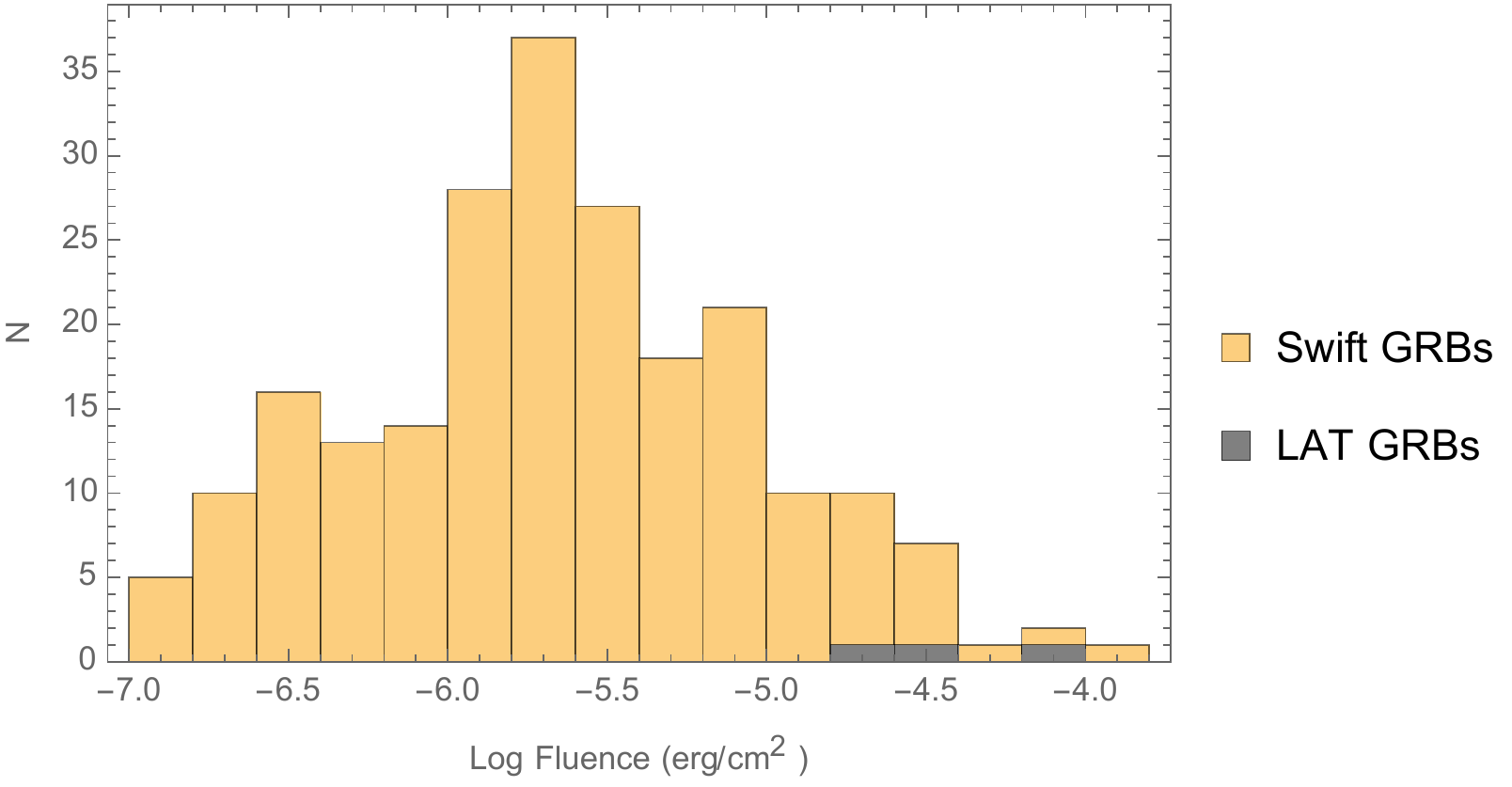}
\caption{Histogram of time $T_X$ of GRBs represented by yellow bins, while $T_\gamma$ is represented by gray bins in the top panel. Bottom panel shows histogram of the fluences, again with \textit{Swift} GRBs in yellow and LAT in gray.}\label{tahisto}
\end{center}
\end{figure}

\begin{center}
\section{$T_X-T_\gamma$ Comparison}
\label{Comparison}
\end{center}
We here explore the existence of the plateau emission in both the $\gamma$ and X-ray energy bands to answer the third question posed in \S \ref{Intro}. For analogy, in order to substantiate the existence of the plateau for these \lat bursts, we fit these LCs with the same W07 function that was used to fit the \textit{Swift} X-ray plateaus. The time at the end of the plateau $T_2$ is represented as $T_{\gamma}$ for LAT GRBs and $T_X$ for \textit{Swift} GRBs hereafter.
As mentioned earlier, there are three LAT GRBs available that appear to show a plateau; however, we do not have coincident observations for these GRBs in X-rays other than for GRB\,090510 \citep{DePasquale2010}. Thus, our comparison is based on an analysis for the sample of GRBs, not on a one-to-one comparison. As seen from the histogram in the top panel of Figure  \ref{tahisto}, the time $T_\gamma$ of the \lat GRBs is on the high end of the  \Swift distribution. We also show that the fluences for the 3 LAT GRBs are accordingly on the higher end when compared to those of \Swift shown in the bottom panel of Figure \ref{tahisto}. Thus, we can suggest that there is a possibility that the end time of the plateau is not achromatic (chromatic), because end point of the plateau is not observed at the same time in $\gamma$-rays and X-rays, as we can see from the differences in $T_{\gamma}$ and $T_X$ in our fitted LCs. In fact, it is worth mentioning that GRB\,090510 has quite different $T_{\gamma}$ and $T_X$ values. 
In case the end time of the plateau is chromatic, this feature is not expected in the energy injection model. We note here that this feature of chromaticity is not present between the X-ray and optical plateaus, as it has been demonstrated in the recent work of \citet{Dainotti2020b}.
This discrepancy between the two wavebands is possibly due to selection effects. Only further observations will allow us to apply meaningful statistical methods to cast further light on whether or not a selection effect is occurring, because there are too few LAT observations in our sample to obtain a statistically significant result. Though the lack of contemporaneous GRBs observed by \lat and XRT prevents us from drawing a definite conclusion, from a statistical point of view, the plateaus seen in X-rays can help us shed light on the differences and similarities between the start and end time of the plateaus in these two different energy ranges.

\FloatBarrier
\FloatBarrier

\section{The 3--D Fundamental Plane Relation}
\label{Results from SWIFT and LAT}

We also check whether the LAT GRBs follow the 3--D fundamental plane relation between the rest frame time at the end of the plateau, peak prompt luminosity, and luminosity at the end of the plateu --
($\log(T_a)$, $\log(L_{peak})$, $\log(L_a)$). This relation is the combination of two relations, one between the luminosity at the end of the plateau emission, $\log(L_{X,a})$ and the end time of the plateau emission, $T_{X,a}$ \citep{Dainotti2008,  dainotti11b, dainotti17}
This relation has been also extended to the optical emission \citep{Dainotti2020b}. This correlation has been interpreted within the magnetar scenario \citep{rowlinson14, rea15, Bernardini2015, Stratta2018}. The second correlation is between the peak luminosity of the prompt emission and the luminosity at the end of plateau emission \citep{dainotti11b, dainotti15}. 
This relation, fitted with our set of 222 \textit{Swift} GRBs with redshift, has been first discovered by \citet{dainotti16c} and later updated in \citet{dainotti17}, \citet{Srinivasaragavan2020} and \citet{Dainotti2020a}. The best fit equation of the plane has the form of :
\begin{equation}
\log L_a= C_o + a \log T_a + b \log L_{peak}
\label{planeequation}
\end{equation} 
where $C_o$ represents the normalization of the plane, and  $a$ and $b$ are  the best fit slope parameters for $T_a$ and $L_a$. The \textit{Swift} data has best-fit parameters of  $C_o$ = 8.53, $a$ = -0.72, and $b$ = 0.81.

\begin{figure}[h!]
\centering 
\includegraphics[width=0.6\linewidth]{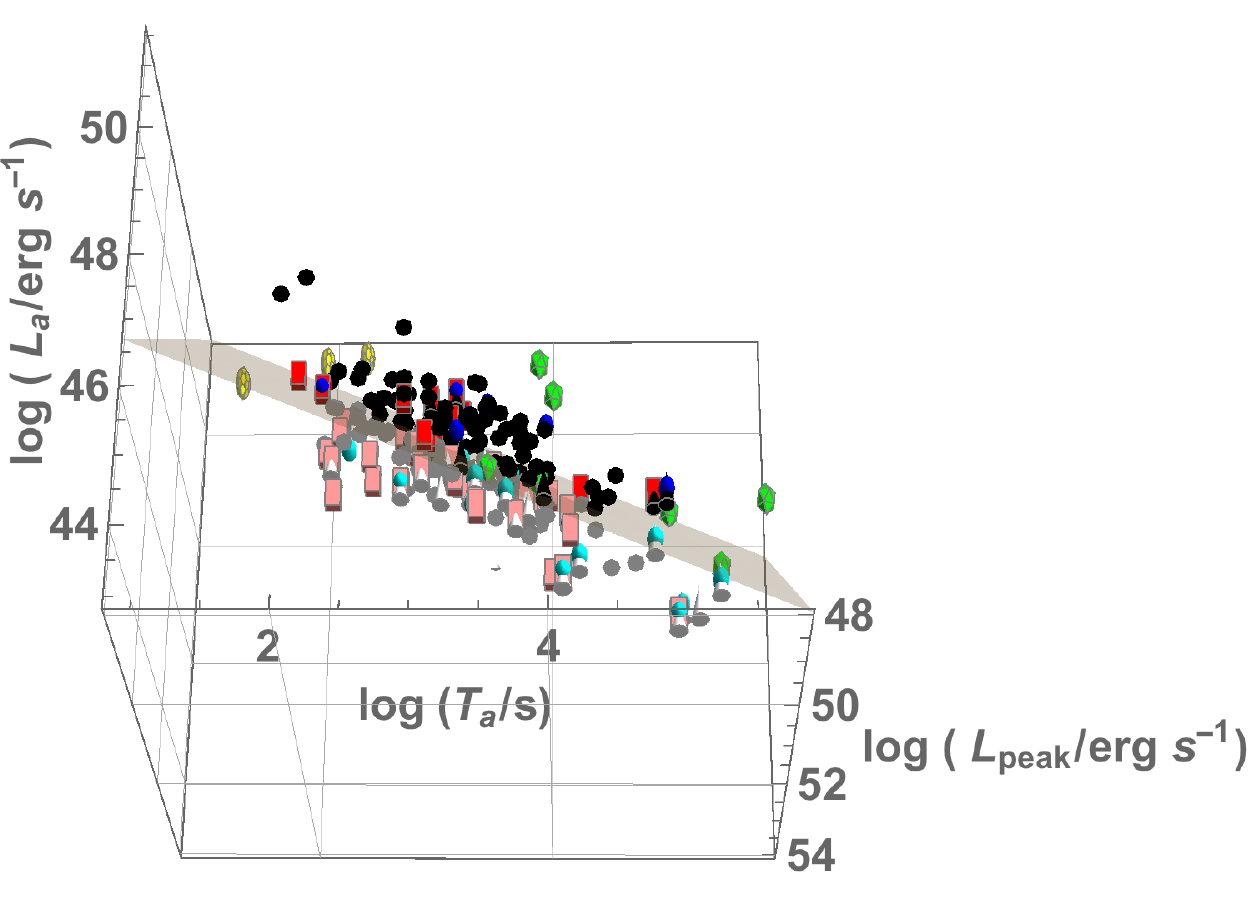}
  \includegraphics[width=0.55\linewidth]{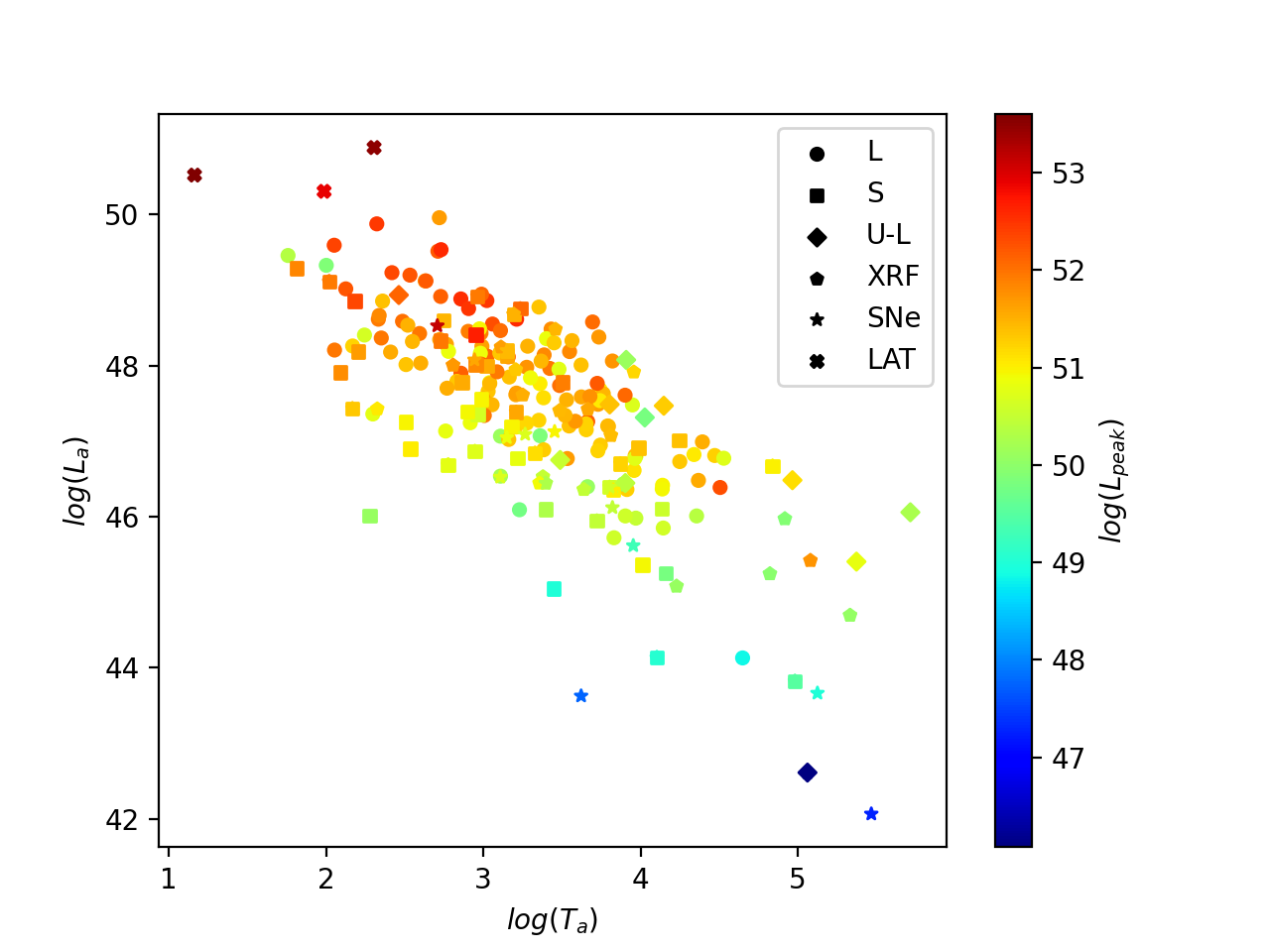}
  \includegraphics[width=0.6\linewidth]{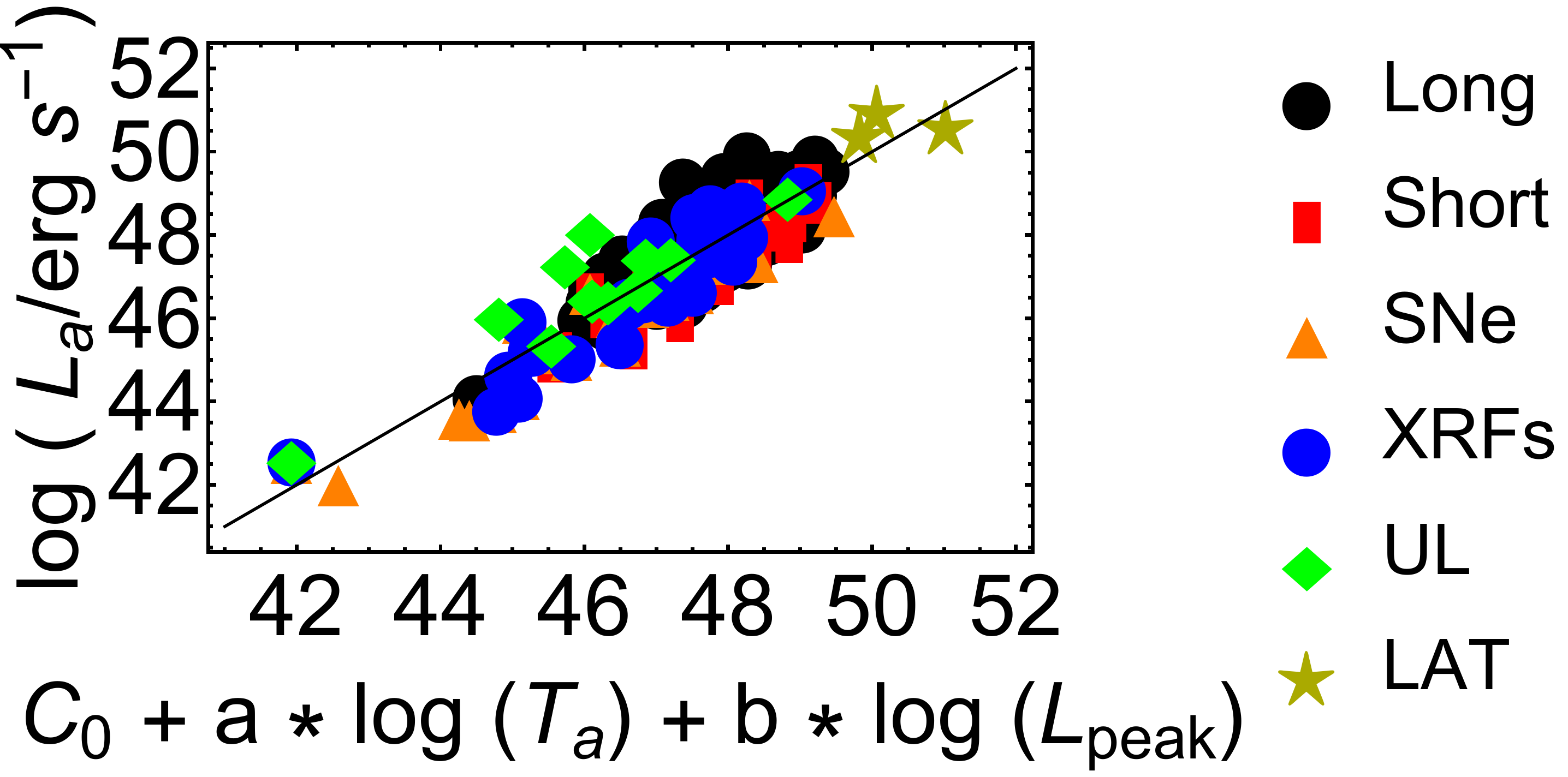}
 \caption{In the upper panel, 222 GRBs in the 3D $T_a-L_{peak}-L_a$ space, with a plane fitted to the data with the following classifications: long GRBs (circles), short GRBs (cuboids), GRB-SNe (cones), XRFs (spheres),  ultra-long GRBs (green icosahedrons), and \Fermi-LAT GRBs (yellow truncated icosahedrons). Darker colors indicate data points above the plane, while lighter colors indicate data points below the plane, except for ultra-long GRBs which are all denoted by bright green icosahedrons. The middle panel shows the same 3D relation divided into subclasses according to the legend, but in which $L_{peak}$ is represented by the color bar.  
The same data is shown in the bottom panel, but represented as a 2--D projection of the 3--D plane with the best-fit line shown in black. }\label{3Dplot}
\end{figure}

We show in the upper panel of Figure \ref{3Dplot} a 3D projection of the correlation with the several categories classified according to \citet{dainotti17},  presented in different shapes and colors:  long  GRBs  (circles),  short GRBs (cuboids), GRBs associated with supernovae (GRB-SNe,  cones), those with X-ray fluence (2 - 30 keV) $>$ $\gamma$- ray fluence (30 - 400 keV) (XRFs, spheres),    ultra-long GRBs with $T_{90} \geq 1000s$ \citep[green polyhedrons]{levan14, Stratta2013, Nakauchi2013},  and  LAT  GRBs  (yellow isocahedrons). Darker  colors  indicate  GRBs above the plane, while lighter colors show GRBs below the plane. For clarity in the middle panel of Figure \ref{3Dplot} we show the 2D projection of the fundamental plane in which the variable $L_{peak}$ is shown with a color bar gradient. In the lower panel of Fig. \ref{3Dplot} we show the 2D projection of the fundamental plane with $L_a$ as a function of $T^{*}_a$ and $L_{peak}$. We note that the three GRBs showing plateaus in $\gamma$-rays obey the 3D correlation observed in X-rays by \textit{Swift}, though their time $T_\gamma$ is on the lower end when compared to GRBs observed by \textit{Swift}. These GRBs are represented by the yellow isocahedrons in the left panel of Figure \ref{3Dplot} and as dark yellow stars in the right panel of Fig. \ref{3Dplot}.  This conclusion encourages us to further pursue this line of research and add more GRBs observed by LAT in a future analysis. This will lend weight to the interpretation of a better fit for the plateau emission than for the PL case as we have shown in the three GRBs studied which present a plateau phase.
\FloatBarrier

\section{Summary and Conclusions} \label{conclusion}
To summarize, we examine the LCs observed by LAT from July 2008 until August 2016 with $TS > 64$ contained in the 2FLGC, selecting the ones that could be reliably fitted within the W07 model to understand if a plateau emission in $\gamma$-rays exists. We test a set of 9 CRs on the GRBs presenting plateaus and with known redshifts and check whether they are fulfilled as a quick validation of the ES model. We also compare the time at the end of the plateaus in $\gamma$-rays and X-rays, through comparing the LAT GRBs with the analysis of 222 GRBs with known redshifts detected by \Swift from January 2005 up to July 2019 that have a plateau, and test the 3--D fundamental plane relation on this set of GRBs in addition to the LAT GRBs we analyse. 
In conclusion, through our analysis of \lat LCs, we have answered the main queries we targeted in Section \ref{Intro}. 
\begin{enumerate}
\item We find three GRBs with known redshifts that show a plateau (Figure  \ref{composite}) similar to the ones found in many X-ray afterglows, which in our opinion highlights the importance of a further study of this point.  
\item 
 The most favorable scenario for the GRBs in our analysis is a constant density ISM or a wind environment with slow cooling, while the least favorable scenario is a constant density ISM or wind environment with fast cooling. Each of the GRBs analysed fulfills at least one CR pertaining to the ES model.  

For GRB\,090510, 2 CRs are fulfilled. When looking at these relations, and also the coincident X-ray LCs, we conclude with the help of the analysis of \citet{Kumar2010} that the most likely scenario for this GRB is a constant ISM with slow cooling.

For GRB\,090902B, 1 CR is  fulfilled. With the help of the analysis of \citet{Ajello2018}, we conclude that the most likely scenario for this GRB is a wind environment with slow cooling.

For GRB\,160509A, 1 CR is fulfilled, and we can conclude that the most likely scenario for this GRB is a wind environment with slow cooling.

We also see that the interpretation of the emission of some GRBs is consistent with existing literature, while for others it is not.



The discrepancy between some of our results and the ones of \citet{Kumar2010} may be due to three ingredients: first, the prompt sub-MeV emission in \citet{Kumar2010} was not considered, so the first few points of the LCs were not fitted. In addition, we here use PASS 8 (which was not available in 2010) for the analysis. PASS 8 provides a better effective area and energy resolution, and consequently allows us to verify the existence of the plateau. Finally, the fitting procedure is also different, since we are using the W07 function, while they use a simple PL fitting. Some of our results are also discrepant with those of \citet{Tak2019}, and they can also be attributed to the differences in our fitting procedures. 
Looking at the individual GRBs themselves, we are able to draw conclusions for all the GRBs analysed.

\item We determine that $T_\gamma < T_X$ through comparing their distributions. This may show an indication of chromaticity of the end time of the plateau in these LCs. The chromaticity at the end of the plateau for the specific case of GRB 090510, the only GRB in our set with multi-wavelength data in the plateau emission available, strengthens this hypothesis.

\item We confirm that the three LAT GRBs do follow the 3--D fundamental plane relation fitted with our enlarged data set of 222 \textit{Swift} LCs compared to past analysis \citep{dainotti16c, dainotti17}.
\end{enumerate}
For all of the above reasons, the further investigation of more high energy light curves so as to cast light on the problems discussed, becomes compelling.  

\section{Acknowledgements}
M.G.D. is particularly grateful to Donald Warren and Hirotaka Ito for the fruitful discussions about Section 4 related to $\epsilon_b$.
M.G.D. is grateful to funding from the European Union FP7 scheme the Marie Curie Outgoing Fellowship, because the research leading to these results have been founded under the contract number N 626267. M.G.D. is also grateful to MINIATURA2 Number 2018/02/X/ST9/03673: and the American Astronomical Society Chretienne Fellowship. M.G.D. is also grateful to be hosted in January and February 2019 by S. Nagataki with the support of "RIKEN Cluster for Pioneering Research". S. Nagataki is grateful to the Pioneering Program of RIKEN for Evolution of Matter in the Universe(r-EMU)"
S. Nagataki also acknowledges the "JSPS Grant-in-Aid for Scientific Research "KAKENHI" (A) with Grant Number JP19H00693." G. Srinivasaragavan is grateful for the support of the United States Department of Energy in funding the Science Undergraduate Laboratory Internship (SULI) program. 
X. Hernandez acknowledges financial assistance from UNAM DGAPA grant, IN106220 and CONACYT.
M. Axelsson gratefully acknowledges funding from the European Union’s Horizon 2020 research and innovation programme under the Marie Sklodowska-Curie grant agreement No 734303 (NEWS). Paul O' Brien acknowledges support from the UK Science and Technology Facilities Council. This work made use of data supplied by the UK Swift Science Data Centre at the University of Leicester.

The \Fermi-LAT Collaboration acknowledges generous ongoing support from a number of agencies and institutes that have supported both the development and the operation of the LAT as well as scientific data analysis. These include the National Aeronautics and Space Administration and the Department of Energy in the United States, the Commissariat \'a l’Energie Atomique and the Centre National de la Recherche Scientifique / Institut National de Physique Nucl\'eaire et de Physique des Particules in France, the Agenzia Spaziale Italiana and the Istituto Nazionale di Fisica Nucl\'eare in Italy, the Ministry of Education, Culture, Sports, Science and Technology (MEXT), High Energy Accelerator Research Organization (KEK) and Japan Aerospace Exploration Agency (JAXA) in Japan, and the K. A. Wallenberg Foundation, the Swedish Research Council and the Swedish National Space Agency in Sweden. Additional support for science analysis during the operations phase is gratefully acknowledged from the Istituto Nazionale di Astrofisica in Italy and the Centre National d’Etudes Spatiales in France. This work was performed in part under DOE Contract DE-AC02-76SF00515.


\bibliography{bib_Fermi_paper_version9.tex}
\bibliographystyle{yahapj}

%

%


\end{document}